\documentclass[12pt]{article}
\usepackage{fullpage,epsfig,graphics,amsbsy,amssymb,cancel,bbm,cite}
\usepackage{psfrag,hyperref,color,slashed}
\usepackage{graphicx}
\usepackage{amsfonts,mathdots,dsfont}
\usepackage{amssymb,amsmath,amscd,mathrsfs}
\usepackage[T1]{fontenc}
\usepackage{tikz}
\usetikzlibrary{arrows,chains,matrix,positioning,scopes,snakes,fadings}

\makeatletter
\tikzset{join/.code=\tikzset{after node path={%
\ifx\tikzchainprevious\pgfutil@empty\else(\tikzchainprevious)%
edge[every join]#1(\tikzchaincurrent)\fi}}}
\makeatother

\tikzset{>=stealth',every on chain/.append style={join},
         every join/.style={->}}
\tikzset{
    >=stealth',
    punkt/.style={
           rectangle,
           rounded corners,
           draw=black, very thick,
           text width=6.5em,
           minimum height=2em,
           text centered},
    pil/.style={
           ->,
           thick,
           shorten <=2pt,
           shorten >=2pt,}
}

\newcommand{\BB}{\mathbb}

\def\alphadot{{\dot{\alpha}}}
\def\betadot{{\dot{\beta}}}

\newcommand{\bea}{\begin{eqnarray}}
\newcommand{\eea}{\end{eqnarray}}
\newcommand{\be}{\begin{equation}}
\newcommand{\ee}{\end{equation}}
\newcommand{\nn}{\nonumber}
\newcommand{\Tr}{\textrm{Tr}}

\newcommand{\sbullet}{\vcenter{\hbox{\tiny$\bullet$}}}

\newcommand{\bra}{\langle}
\newcommand{\ket}{\rangle}

\newcommand{\ind}{\textrm{ind}}

\def\gdh{\textrm{\dh}}
\newcommand{\sdet}{\operatorname{sdet}}

\def\ga{\alpha}

\def\gd{\delta}
\def\Gd{\Delta}
\def\ep{\epsilon}

\def\ep{\epsilon}

\def\gl{\lambda}

\def\Go{\Omega}
\def\go{\omega}

\DeclareMathAlphabet{\mathpzc}{OT1}{pzc}{m}{it}

\usepackage{amsthm}
\usepackage{amsmath}
\usepackage{amsfonts}
\usepackage{amssymb}

\theoremstyle{definition}

\begin{document}
\begin{flushright} \small
UUITP-59/18
 \end{flushright}
\smallskip
\begin{center} \Large
{\bf  Twisting with a Flip\\
(the Art of Pestunization)}
 \\[12mm] \normalsize
{\bf Guido Festuccia${}^a$, Jian Qiu${}^{a,b}$, Jacob Winding${}^c$, Maxim Zabzine${}^a$} \\[8mm]
 {\small\it
   ${}^a$Department of Physics and Astronomy,
     Uppsala University,\\
     Box 516,
     SE-75120 Uppsala,
     Sweden\\
   \vspace{.5cm}
      ${}^b$ Mathematics Institute,  Uppsala University, \\
   Box 480, SE-75106 Uppsala, Sweden\\
    \vspace{.5cm}
    ${}^c$
    School of Physics, Korea Institute for Advanced Study,\\
     Seoul 130-722, Korea\\
   }
\end{center}
 \vspace{7mm}
\begin{abstract}
	\noindent	We construct ${\cal N}=2$ supersymmetric Yang-Mills theory on 4D manifolds with a Killing vector field with isolated fixed points.
	It turns out that for every fixed point one can allocate either instanton or anti-instanton contributions to the partition function,
	and that this is compatible with supersymmetry. The equivariant Donaldson-Witten theory is a special case of our construction.
	We present a unified treatment of Pestun's calculation on $S^4$ and  equivariant Donaldson-Witten theory
	by generalizing the notion of self-duality on manifolds with a vector field.
	We conjecture the full partition function for a
	${\cal N}=2$ theory on any 4D  manifold with a Killing vector. Using this new notion of self-duality to localize a supersymmetric theory is what we call ``Pestunization''.
	\noindent
\end{abstract}
\eject
\normalsize

\tableofcontents
\section{Introduction}

Starting from the work \cite{Pestun:2007rz} the localization of supersymmetric theories on compact manifolds has attracted considerable attention (see \cite{Pestun:2016zxk} for a review of the latest developments).
Supersymmetric theories were constructed on diverse 2D to 7D compact manifolds and their partition functions were calculated or conjectured.
Typically the manifold admits (generalized) Killing spinors and there is enough torus action to proceed with the localization calculation.
In low dimensions toric geometry is not very rich; it becomes interesting starting from 4D and higher.
Constructing supersymmetric field theories on odd dimensional manifolds $M$ is simple. This is related to the fact that Killing spinors on $M$ are related to covariantly constant spinors on the cone over $M$. For example, any toric Calabi-Yau cone in 6D produces a toric 5D Sasaki-Einstein geometry (i.e. a geometry with two Killing spinors). A similar story holds in 7D. In even dimensions the situation is more complicated.
In this paper we concentrate on ${\cal N}=2$ supersymmetric theories on 4D compact manifolds with a $T^2$-action.
Our goal is to explain their structure and also to show how the localization result for their partition function is organized.

Let us briefly review different ways of placing ${\cal N}=2$ theories on 4D manifolds preserving some supersymmetry.
About 30 years ago Witten \cite{Witten:1988ze} constructed Donaldson-Witten theory, which corresponds to a topological twist of ${\cal N}=2$ supersymmetric gauge theory and  localizes on instantons.
Donaldson-Witten theory is related to the calculation of Donaldson invariants \cite{MR1066174, donaldson1990geometry}.
If the 4D manifold admits a torus action then one can define equivariant Donaldson-Witten theory.
This theory has been studied in detail on $\mathbb{R}^4$  \cite{Losev:1997tp, Moore:1997dj, Lossev:1997bz, Moore:1998et} and the corresponding partition function is known as the Nekrasov partition function $Z^{\rm inst}_{\epsilon_1, \epsilon_2} (a, q)$ \cite{Nekrasov:2002qd, Nekrasov:2003rj}.
Schematically the Nekrasov function is
\bea
Z^{\rm inst}_{\epsilon_1, \epsilon_2} (a, q) = Z^{\rm 1-loop}(a) \sum\limits_{n=0}^\infty q^n ~ {\rm vol}_n(\epsilon_1, \epsilon_2, a) ,
\eea
where ${\rm vol}_n(\epsilon_1, \epsilon_2, a)$ is the equivariant volume of the moduli space of instantons of charge $n$ and we include the perturbative 1-loop contribution $ Z^{\rm 1-loop}(a)$ for later convenience.
The partition function of equivariant Donaldson-Witten theory on non-compact toric surfaces has been later conjectured in \cite{MR2227881} (see also \cite{Gottsche:2006tn, Gottsche:2006bm, Gasparim:2008ri} for further studies and proofs).
The full answer is obtained gluing copies of Nekrasov partition functions associated to each fixed point of the torus action, with some discrete shifts associated to fluxes (related to non-trivial second cohomology).
A second class of supersymmetric theories originates from the work of Pestun \cite{Pestun:2007rz} who placed a ${\cal N}=2$ supersymmetric gauge theory on the round $S^4$. This construction was later extended to certain squashed spheres $S_{\epsilon_1, \epsilon_2}^4$ in \cite{Hama:2012bg, Pestun:2014mja}. The resulting theory is not related in any obvious way to equivariant Donaldson-Witten theory.
Their partition function on $S_{\epsilon_1, \epsilon_2}^4$ can be written schematically as follows
\bea
Z_{S_{\epsilon_1, \epsilon_2}^4} = \int da~e^{S_{cl}} ~Z^{\rm inst}_{\epsilon_1, \epsilon_2} (ia, q)   Z^{\rm anti-inst}_{\epsilon_1, - \epsilon_2} (ia, \bar{q})~,\label{Pestun-result}
\eea
displaying the contribution of instantons over the north pole and anti-instantons over the south pole of $S^4$ (the way we write the expression for $Z_{S_{\epsilon_1, \epsilon_2}^4} $  is clarified in section \ref{ss:full-answer}).

In  \cite{Festuccia:2016gul}, reducing a 5D theory,  a ${\cal N}=2$ theory was constructed on  the connected sum $\#_k (S^2 \times S^2)$, which is a toric manifold with $T^2$-action, and has $(2+2k)$ fixed points.
Similarly to (\ref{Pestun-result}) the partition function\footnote{ In conjecturing the answer for the partition function in \cite{Festuccia:2016gul} we missed the contribution of fluxes.}
for $\#_k (S^2 \times S^2)$ is obtained gluing $(k+1)$ copies of  Nekrasov functions for instantons and $(k+1)$-copies for anti-instantons.

From the aforementioned results a question arises naturally. Consider any simply connected compact 4D manifold with a $T^2$-action with isolated fixed points. Distribute Nekrasov partition functions for instantons over some of the fixed points and partition functions for anti-instantons over the rest. Does this correspond to the partition function of some supersymmetric theory? In this paper we answer  this question positively. The case when we associate instantons to all fixed points corresponds to equivariant Donaldson invariants,  (see e.g. \cite{Bershtein:2015xfa, Bershtein:2016mxz} and earlier works  \cite{Bawane:2014uka, Sinamuli:2014lma, Rodriguez-Gomez:2014eza}).
We present a general class of supersymmetric field theories that includes on equal footing both equivariant Donaldson-Witten theory and the theories on $S_{\epsilon_1, \epsilon_2}^4$ reviewed above. In the rest of this section we briefly explain the main ideas leading to our construction.

Donaldson-Witten theory is a 4D gauge theory counting instantons ({anti} self-dual connections) in some appropriate sense.
{Anti} self-duality is a nice problem in 4D and its linearization is related to the following elliptic complex on 4D manifolds
\bea
\Omega^0 (M_4)~\xrightarrow{d}~\Omega^1(M_4)~\xrightarrow{P^+ d}~P^+\Omega^{2}(M_4) = \Omega^{2+}(M_4)~,\label{elliptic-complex}
\eea
where $\Omega^p$ are the $p$-forms, $d$ is de Rham differential and  the projectors $P^\pm = \frac{1}{2} (1\pm \star)$ use the Hodge star $\star$.
Ellipticity of the complex (\ref{elliptic-complex}) (or ellipticity of the corresponding PDEs) is closely related to the existence of a finite dimensional moduli space of instantons.
Using the projector $P^+$ the Yang-Mills action can be written as follows
\bea
||F||^2 = ||P^+ F||^2 + ... ~, \label{YM-instanton}
\eea
where the dots stand for the topological term.
Donaldson-Witten field theory localizes to $P^+ F=0$, and provides an infinite dimensional analogue of the Euler class of a vector bundle \cite{Atiyah:1990tm}. Eventually the calculation is reduced to a finite dimensional problem on the moduli space of instantons.
Most of the known cohomological and topological field theories are related to underlying finite dimensional moduli spaces, and thus to some elliptic problem.

In \cite{Pestun:2007rz} Pestun uses the theory of transversely elliptic operators (correspondingly there are notions of transversely elliptic complex and transversely elliptic PDEs) in order to calculate the 1-loop determinants arising from localization.
It is one of the goals of this paper (and its follow up \cite{Fest-2018}) to show that supersymmetry is closely related to such transversely elliptic problems.

In order to introduce the logic underlying our later constructions, we provide below an informal explanation of the notion of transverse ellipticity.
If on a compact manifold $M$ we have an action of a group $G$ then we say that an operator is transversely elliptic if it is elliptic in the directions normal to the $G$-orbits.
In this case the kernel and co-kernel of this operator are not finite dimensional, but they can be decomposed in finite dimensional representations of $G$.
As an example consider the elliptic problem in 2D corresponding to the Dolbeault operator $\bar{\partial}$ acting on functions\footnote{Strictly speaking $\bar{\partial}$ sends functions
 to sections of the anti-canonical line bundle. For the introduction we ignore this distinction. Similar remark applies to $\bar{\partial}_H$ on  $S^1 \times S^2$ and $S^3$ below.}, e.g.  functions $\phi$ on $S^2$ satisfying the condition $\bar{\partial} \phi=0$.
If we add a circle $S^1$ to this problem and consider a new function $\phi$ on  $S^1 \times S^2$ satisfying the condition  $\bar{\partial} \phi=0$, we obtain an example of a 3D transversely elliptic problem with respect to the $S^1$-action. There is no finite dimensional kernel for this 3D problem; however, we can decompose $\phi$ according to ${\cal L}_v\phi_n = i n \phi_n$ with $v$ being a vector field along the to $S^1$ (i.e. we decompose $\phi$ into Fourier modes).
The problem $\bar{\partial} \phi_n =0$ has a finite dimensional kernel/co-kernel.
This setup can be covariantized and we can consider a function $\phi$ on  $S^3$ with the $S^1$-action associated to the Hopf fibration and where $\bar{\partial}_H$ is the corresponding  horizontal Dolbeault operator on $S^3$.
Again the problem $\bar{\partial}_H \phi=0$ is an example of a  transversely elliptic problem with respect to the $S^1$-action, and this equation behaves nicely under ${\cal L}_v$ (in the same way as we just discussed, modulo some technicalities).
It is crucial that the first order transversely elliptic operator gives rise to the standard second order elliptic operator
\bea
\Delta = - {\cal L}_v^2 + \partial_H \bar{\partial}_H~,\label{decom-intro}
\eea
which is just the Laplace operator in this example.
There is also a converse statement: if we take a second order elliptic operator and find a decomposition like \eqref{decom-intro} then the corresponding first order operator $\bar{\partial}$ will automatically be transversely elliptic.
Supersymmetry typically comes with some Killing vector field $v$ and it naturally produces a decomposition of a second order elliptic operator as in \eqref{decom-intro}.
This is a salient feature of many of the supersymmetric models on curved manifolds that are discussed in the context of supersymmetric localization.

Following this informal introduction to transversely elliptic operators, we reconsider 4D gauge theories on a four manifold $M_4$.
As we have mentioned, in this context instantons are related to the natural elliptic problem (\ref{elliptic-complex}).
We can repeat the same trick as above and consider the 5D manifold $M_5 = S^1 \times M_4$ with trivial $U(1)$-action along $S^1$.
There we can construct a 5D transversely elliptic problem by requiring anti self-duality in the transverse directions (along $M_4$) and that the component of the gauge field along $S^1$ is zero.
One can rewrite this system in 5D covariant terms, see equation (\ref{cov-tre-5D})  in appendix \ref{App-TEP}.
This can be summarized in terms of the following transversely elliptic complex
\bea
\Omega^0 (M_5)~\xrightarrow{d}~\Omega^1(M_5)~\xrightarrow{D}~\Omega^{2+}_H(M_5) \oplus \Omega^{0}(M_5)~,\label{5D-tr.elliptic-complex}
\eea
where $H$ stands for horizontal.
At the level of this discussion the concrete form of the operator $D$ is not important, what is crucial is that the 4D instanton equation admits a natural transversely elliptic 5D lift.
 In general this complex is defined for $M_5$ with torus action and reducing along a free $S^1$ will give us some $M_4$.
This reduction will naturally produce a new transversely elliptic problem in 4D with respect to the remaining action of a vector field $v$.
Upon the reduction the complex \eqref{5D-tr.elliptic-complex} becomes symbolically
\bea
\Omega^0 (M_4) ~\xrightarrow{d}~\Omega^1 (M_4) \oplus \Omega^0(M_4)~\xrightarrow{\tilde{D}}~P^+_\omega\Omega^{2} (M_4)\oplus \Omega^0 (M_4)~,\label{new-compex}
\eea
where $P^+_\omega$ is some projector which defines a sub-bundle of rank $3$ within the two forms  $\Omega^{2}$, and the system involves the 4D gauge field and a scalar field (the $\Omega^0(M_4)$-part in the middle term).
The concrete form of the operator $\tilde D$ can be found in appendix \ref{App-TEP} and will be discussed later.
As we will see ${\cal N}=2$ supersymmetry on $M_4$ naturally selects the corresponding transversely elliptic problem (\ref{new-compex}).
The analog of the decomposition (\ref{YM-instanton}) for the Yang-Mills action is
\bea
||F||^2 = || h P_\omega^+ F||^2 + || f \iota_v F||^2 + ... ~,
\eea
where the dots stand for lower derivative terms and $f$, $h$ are some positive functions.
We will refer to the condition   $P^+_\go F = 0$ as defining a \emph{flip instanton}.
In this paper we will derive explicitly the form of the projector $P^+_\go$,  from geometric considerations and show its relation to supersymmetry.  The idea is that the relevant second order elliptic operator can be decomposed in terms of first order operators as in (\ref{decom-intro})  (for a gauge theory this is a subtle statement, but it is roughly correct).
We have provided the 5D perspective as a motivation, but our construction can be defined in intrinsically 4D terms, and supersymmetry is naturally related to the transversely elliptic complex (\ref{5D-tr.elliptic-complex}).

In relating to the original Donaldson-Witten cohomological theory, the novelty here is that on 4D manifolds with a $T^2$ action, the elliptic complex \eqref{elliptic-complex} can be replaced by another complex, which instead is transversally elliptic with respect to the $T^2$ action.
Not surprisingly this new complex, and in particular the new bundle $P^+_\go \Omega^2$, is  naturally related to supersymmetry and the localization calculation to be carried out in this new framework.
If we think about Pestun's $S^4$ construction, the bundle $P^+_\omega\Omega^{2}$ can be thought of as that of self-dual forms over the north hemisphere and of anti-selfdual forms over the south hemisphere. These two spaces are glued together using the vector field coming from the $T^2$-action.
The flip from self-duality to anti self-duality between the two hemispheres motivates why we call solutions to $P^+_\go F = 0$ flip instantons.

\subsection{Summary of results}

We present two main results: the first is the explicit construction of a ${\cal N}=2$ supersymmetric gauge theory on any manifold with a Killing vector field with isolated fixed points.
The second result is a conjecture for the full partition function for these theories.
We outline some technical problems related to proving this conjecture.

Consider a Riemannian manifold $(M,g)$ with a Killing vector field $v$ with isolated fixed points.
Let us choose a decomposition $||v||^2 = s \tilde{s}$ in terms of two non-negative invariant functions $s$ and $\tilde{s}$ such that $\tilde{s}=0$ at some fixed points (we call them {\it plus} fixed points) and $s=0$ at all remaining fixed points (we call them {\it minus} fixed points).
For such data we can construct generalized Killing spinors and the corresponding  ${\cal N}=2$ supersymmetric gauge theory for a vector multiplet.
This geometrical data does not uniquely fix all auxiliary supergravity fields; however, this redundancy does not affect the partition function.
Using the Killing spinors we can reformulate the ${\cal N}=2$ vector multiplet in terms of cohomological variables. These are very similar to those that are used to formulate equivariant Donaldson-Witten theory except that we change the notion of self-duality and correspondingly the definition of the two-form fields (denoted $\chi$ and $H$ below).
We construct a rank 3 subbundle $P^+_\omega\Omega^{2}$ of two forms that near the {\it plus} fixed points approach self-dual two forms, and near the {\it minus} fixed approach anti-self-dual two forms.
Away from the fixed points we use the vector field $v$ to glue self-dual with anti-self-dual forms.
Thus we define a new version of equivariant Donaldson-Witten theory and relate it to supersymmetry.

The second result of our paper is to perform the localization calculation for the partition function for the  ${\cal N}=2$ supersymmetric gauge theory we constructed.
We are not able to derive the form of the partition function in the most general setting. However, we conjecture that two types of contributions appear in the path integral:  point like instantons (at plus fixed points), point like anti-instantons (at minus fixed points), as well as flux configurations related to non-trivial two-cycles on $M$. Here we consider only vector multiplets. Assuming that we have a $T^2$-action on $M$  given by a vector field $v= \epsilon_1 v_1 +  \epsilon_2 v_2$ the full answer can be written schematically as follows
\be
Z_{M_{\epsilon_1, \epsilon_2}} = \!\!\sum\limits_{{\rm\small{discrete~} k_i}}\,\int\limits_{\bf h} \! da\, e^{-S_{cl}} \prod\limits_{i=1}^{p} Z^{\rm inst}_{\epsilon_1^i, \epsilon_2^i} \Big (ia + k_i (\epsilon_1^i, \epsilon_2^i), q \Big )  \prod\limits_{i=p+1}^l  Z^{\rm anti-inst}_{\epsilon_1^i, \epsilon_2^i} \Big (ia + k_i  (\epsilon_1^i, \epsilon_2^i), \bar{q} \Big ) ~,
 \label{full-answer-conj}
\ee
where we have $p$ copies of the Nekrasov instanton partition function $Z^{\rm inst}_{\epsilon_1, \epsilon_2}(ia, q)$  for the $p$ plus fixed points and $(l-p)$ copies of the Nekrasov anti-instanton partition function $Z^{\rm anti-inst}_{\epsilon_1, \epsilon_2}(ia, \bar{q})$ for the $(l-p)$ minus fixed points. The equivariant parameters  $(\epsilon_1^i, \epsilon_2^i)$ can be read off from the local action of $v$ in the neighbourhood of the fixed points.
In  (\ref{full-answer-conj}) $k_i(\epsilon_1^i, \epsilon_2^i)$ are vector-valued linear functions in  $(\epsilon_1^i, \epsilon_2^i)$ with integer coefficients that correspond to the fluxes.  At the moment we are unable to characterize these functions in the general case and we illustrate some problems related to the presence of fluxes. Looking at some examples we show indications that the contribution of fluxes depends on the relative distribution of plus and minus fixed points.
Assuming that $M$ is simply connected we can express the perturbative contribution (around the zero connection) as a formal superdeterminant that can be calculated using index theorems for the transversely elliptic complex (\ref{new-compex}). In general it is not clear how to consistently glue the local contributions to produce a well-defined analytical answer.
Nevertheless, we give some  examples where the perturbative answer can be written explicitly. Finally we argue that the full answer (\ref{full-answer-conj}) depends holomorphically from $\epsilon_1$ and $\epsilon_2$, and that this is compatible with Pestun's answer on $S^4$ for real $\epsilon_1, \epsilon_2$.

\subsection{Outline of the paper}

We summarize the content of each section of the paper. Many formal aspects of our construction are only mentioned in this work and will be discussed in the follow up \cite{Fest-2018}.

In section \ref{s:coh-complex} we  start with the definition of  our cohomological field theory axiomatically.
Our main goal is to define a new decomposition, into two orthogonal subbundles of rank 3, of the two forms $\Omega^2$ on a 4D manifold equipped with a vector field $v$.
We do this both using the language of transition functions and by constructing the projector $P^+_\omega$ explicitly.
Using this new decomposition of two forms we define a cohomological field theory, which is a generalization of equivariant Donaldson-Witten theory.
We briefly discuss the cohomological observables in this theory.

Section \ref{s:supersymmetry} provides the detailed construction of a ${\cal N}=2$ supersymmetric gauge theory on a compact manifold equipped with a Killing vector field $v$ with isolated fixed points. The corresponding Killing spinors are analysed and the Lagrangian for a ${\cal N}=2$ vector multiplet is shown. Local aspects of this construction are not novel (see e.g. \cite{Klare:2013dka,Pestun:2014mja}); however, here we establish that the supersymmetric theory is globally well defined. We provide an explicit map between the ${\cal N}=2$ supersymmetric Yang-Mills theory and the cohomological field theory defined in section \ref{s:coh-complex}.
We show how the projector $P^+_\omega$ arises from supersymmetry.

In section \ref{secdefs} we consider the dependence of supersymmetric observables (e.g.~the partition function) on deformations of the geometrical and non-geometrical data entering the construction of the theory. The partition function is shown to be independent of many of these deformations. We also argue that the partition function depends holomorphically on {\it squashing} parameters.

In section \ref{s:localization} we outline the localization calculation.  In the general case we are not able to carry it out completely. We concentrate on concrete examples and outline the technical problems that arise in the general situation. We discuss both perturbative and non-perturbative parts of the answer. Finally we conjecture the general form of the localization result for the partition function.

In section \ref{s:summary} we summarize the paper, point out open problems and we also provide a short outline of our follow up work \cite{Fest-2018}. The paper has many appendices where we collect a summary of our conventions and a number of technical results.

\section{Cohomological theory}\label{s:coh-complex}

In this section we define the  4D cohomological field theory in axiomatic fashion and in the next section we will explain its relation to supersymmetry.
This cohomological theory is a generalization of equivariant Donaldson-Witten theory where we modify the notion of self-duality.
We consider a 4D manifold with metric $g$ and Killing vector $v$ with isolated fixed points. We construct a novel subbundle of two forms $\Omega^2(M)$ that, in the neighbourhood of a fixed point, look like either self-dual or anti-selfdual two forms.
We use the vector field $v$ to glue self-dual and anti-self-dual forms in one rank 3  bundle.  In the next sub-section we provide the explicit construction of this bundle and later describe the associated cohomological field theory.

\subsection{Decomposition of two forms}\label{s:subs-projector}

Before discussing the field theory we introduce a new decomposition of two forms $\Omega^2$ on a compact 4D manifold $M$.
Consider the metric $g$ on $M$ and the corresponding Hodge star $\star$.
On the space of two forms $\Omega^2$ we have the scalar product
\bea
\langle B_1, B_2   \rangle = \int\limits_{M} B_1 \wedge \star B_2~, \qquad B_1,B_2 \in \Omega^2(M)~.
\eea
Thus upon choosing the orientation, the bundle $\Omega^2$ has the structure group $SO(6)$.
Using the Hodge star $\star$ we can introduce projectors $P^\pm = \frac{1}{2} (1\pm \star)$ and decompose $\Omega^2 = \Omega^{2+} \oplus \Omega^{2-}$ into two orthogonal sub-bundles of rank $3$.
This decomposition depends only on the conformal class of the metric $g$.
Assume additionally that there is a vector field $v$ which is nowhere zero (for the moment) and define its dual one form $\kappa= g(v)$.
We can define a map $ m: \Omega^{2+} \rightarrow \Omega^{2-}$,
\bea
m :~ B~\mapsto~ -B + \frac{2}{\iota_v \kappa} \kappa \wedge \iota_v B~,\label{map-m-def}
\eea
where $\iota_v \kappa = g(v,v) = ||v||^2$. To see that the map (\ref{map-m-def}) sends self-dual forms $\Omega^{2+}$ to anti-selfdual $\Omega^{2-}$ forms and vice versa one has to use the following identity on two forms
\bea
\iota_v  (\kappa \wedge  \star B ) =  \star (\kappa \wedge \iota_v B)~.
\eea
The map $m$ satisfies the properties $m^2 =1$ and $m \star + \star m=0$.
Additionally it preserves the scalar product,
\bea
\langle m(B_1), m(B_2) \rangle = \langle B_1, B_2 \rangle~.
\eea
Next we relax the assumptions on the vector field and let $v$ have isolated zeros on $M$.
In this case the map (\ref{map-m-def}) is not globally well defined (it is singular where $v=0$); however, we can use it as a transition function to glue self-dual and anti self-dual forms and define a new bundle.
Let us be more precise. Assume that the vector field $v$ is defined everywhere and that it has a finite number of isolated zeros. Without loss of generality  we can choose an open covering of the manifold, $M =\cup~ U_{i}$, such that every open set contains exactly one zero of $v$ and the double intersections $U_i \cap U_j$ do not contain any zeros.
Next for each zero of $v$ we choose a sign $+$ or $-$  and assign the same sign to the open set that contains the given zero.
We denote the open sets correspondingly as $U^+_i$ and $U^-_j$.
We associate the self-dual forms $\Omega^{2+}(U^+_i)$ to the sets $U^+_i$ and anti self-dual forms $\Omega^{2-}(U^-_j)$ to the sets $U^-_j$. Over the intersections $U^+_i \cap U^-_j \neq \emptyset$ we use the transition function given by the map $m_{ij}: \Omega^{2+}(U^+_i) \rightarrow \Omega^{2-}(U^-_j)$ defined by (\ref{map-m-def}). We glue patches with the same sign through the identity map. The maps $m_{ij}$ satisfy the cocycle condition, thus this gluing defines a rank $3$ subbundle of $\Omega^2(M)$, which we denote throughout the paper as $P^+_\omega\Omega^{2}$.
We will need a more concrete description of this subbundle, hence we will provide an alternative definition using a projector operator.
We will see in the following that this projector is naturally related to supersymmetry.

The projector can be derived in different ways.
We first present an abstract derivation which is directly related to the above formal construction.
For the sake of clarity let us consider the simple situation when manifold is covered by two patches $M=U^+ \cup U^-$  and thus the vector field $v$ has only two zeros.
Assume that we have a partition of unity for this covering, $\phi^+ + \phi^-=1$ and ${\rm supp} (\phi^\pm) \subseteq U^\pm$.
Further assume that our rank 3 subbundle of $\Omega^{2}$ is defined by some projector $P_\omega^+$. Then on the corresponding patches we have
\bea
P_\omega^+ = W^{-1} P^+ W~~~~{\rm on}~~ U^+~,~~~~~~  P_\omega^+ = V^{-1} P^- V~~~~{\rm on}~~ U^-~,
\eea
where $W$, $V$ are special orthogonal transformations on $U^+$ and $U^-$ respectively, and $P^\pm = \frac{1}{2} (1\pm \star)$.
Thus on the intersection $U^+ \cap U^-$ we have the relation
\bea
P^+ = (V W^{-1})^{-1} P^- V W^{-1}~,\label{on-intersect}
\eea
which encodes the gluing map (\ref{map-m-def}).  To be concrete we will use $2$ by $2$ block matrices corresponding to the splitting
$\Omega^2 = \Omega^{2+} \oplus \Omega^{2-}$.  For example, the projector $P^+$ has the form
\bea
P^+ = \bigg (
\begin{array}{cc}
	1 & 0 \\
	0 & 0 \\
\end{array}\bigg )~.
\eea
Using these notations the relation (\ref{on-intersect}) can be written as follows
\bea
V W^{-1} = \bigg (
\begin{array}{cc}
	0 & m \\
	-m & 0 \\
\end{array}\bigg )~,
\eea
where $m$ is the map (\ref{map-m-def}) and the minus sign guarantees that this matrix is an element of $SO(6)$.
We can represent  $W$ and $V$ as elements of $SO(6)$,
\bea
W (\rho)= \bigg (
\begin{array}{cc}
	\cos \rho & m \sin \rho \\
	-m \sin \rho  &  \cos \rho \\
\end{array}\bigg )
~, ~~~~
V (\rho)= \bigg (
\begin{array}{cc}
	- \sin \rho & m \cos \rho \\
	-m \cos \rho  & - \sin \rho \\
\end{array}\bigg ) = W(\rho + \frac{\pi}{2})~,
\eea
where $\rho$ is defined through the partition of unity as follows:  $\phi^+ = \cos \rho$ and $\phi^-= 1- \cos \rho$.
Since $\phi^\pm \geq 0$, we have $\rho \in [-\pi / 2, \pi /2 ]$.
Thus around the zero of $v$ in $U^+$, $\rho$ approaches zero and around the zero of $v$ in $U^-$, $\rho$ approaches $-\pi /2$ (or $\pi / 2$ depending on conventions).
Hence we obtain for $P_\omega^+$
\bea
P_\omega^+ =  W^{-1} P^+ W =   \bigg (
\begin{array}{cc}
	\cos \rho & - m \sin \rho \\
	m \sin \rho  &  \cos \rho \\
\end{array}\bigg )
\bigg (
\begin{array}{cc}
	1 & 0 \\
	0   &   0 \\
\end{array}\bigg )
\bigg (
\begin{array}{cc}
	\cos \rho & m \sin \rho \\
	-m \sin \rho  &  \cos \rho \\
\end{array}\bigg )~,
\eea
and using our conventions we rewrite it as follows
\bea
P_\omega^+ = \cos^2 \rho~ P^+    +   \sin^2 \rho ~P^- +  \cos \rho \sin \rho ~m = \frac{1}{2} ( 1 + \cos 2\rho ~\star + \sin 2\rho ~m)~,\label{new-projector-1}
\eea
where $2\rho \in [ - \pi, \pi]$.
The projector is well defined even at $v=0$ since at those points $\sin 2\rho$ is zero. It is straightforward to generalize this construction to the case where $v$ has more than two zeros and to any allocation of signs to the fixed points.

Indeed we can give the following direct construction of the projector $P_\omega^+$. Away from the zeros of $v$ the following identities hold on $\Omega^2$: $\star^2=1~$, $m^2=1~,$ and $~\star\, m + m\, \star =0~.$ Hence, provided that  $\alpha^2 + \beta^2=1$ the combination $\alpha \star + \beta m$ satisfies
\bea
(\alpha \star + \beta m)^2 =1~.
\eea
It follows that (\ref{new-projector-1}) is the most general projector composed from $\star$ and the vector field. We have to ensure that it can be extended to the zeros of $v$ and thus be well-defined over the whole manifold. For this we need that $\sin 2 \rho$ goes to zero (at least linearly) where  $v=0$. Hence at the fixed points $2 \rho$ goes to either $0$ or $\pi$ implying that  $\cos 2\rho$ goes to $\pm 1$ respectively, and the projector $P_\omega^+$ approaches either  $\frac{1}{2} (1+ \star)$ or $\frac{1}{2} (1-\star)$. It is not hard to construct a function $\rho$ with the required properties and in appendix \ref{app:explicitgeom} we give some explicit examples.

There is some redundancy in our description: as $\cos 2\rho$ changes from $+1$ to $-1$, $2\rho$ may go either from $0$ to $\pi$ or from $0$ to $-\pi$. We fix this ambiguity assuming that $2\rho \in [0, -\pi]$.
This allows us to perform the following change of variables
\bea
1- \sin 2\rho = \frac{2}{1+ \cos^2 \omega}~,
\eea
using a function $\omega(x)$.  The projector (\ref{new-projector-1}) can then be expressed as follows
\bea
\label{projector}
P^+_\omega ={1\over 1+\cos^2\!\omega}\left( 1+  \cos\omega \star-\sin^2\!\omega\,{\kappa\wedge\iota_v \over \iota_v \kappa } \right)~.
\eea
Here the function $\omega$ is chosen in such a way that the last term is well-defined at $v=0$. Moreover $\omega=0$ at the $+$ fixed points and  $\omega=\pi$ at the $-$ fixed points. This reparametrization of the projector is better suited for the considerations in the next section. This projector depends only on the conformal class of the metric. Thus in a given conformal class one can choose a representative for which $||v||^2 = \sin ^2 \omega$ (i.e., $||v||^2 \leq 1$) and the projector is somewhat simpler,
\bea
\label{projector-simpler}
P^+_\omega ={1\over 2- ||v||^2}\left( 1+ \sqrt{1-||v||^2}  \star- \kappa\wedge\iota_v  \right)~,
\eea
(Note however that $v$ does not define $\omega(x)$ uniquely because of the ambiguity in taking the square root).
Some formulas below are simpler for a metric with this special property and the corresponding projector (\ref{projector-simpler}), but everything we present holds for a generic choice of metric.

Coming back to the general projector (\ref{projector}) and using the identities presented in this section we write the following useful formula for a two form $F$
\bea
(1+\cos^2 \omega)  P^+_\omega F \wedge \star  P^+_\omega F  =  F\wedge \star F + \cos \omega~ F \wedge F - \frac{\sin^2 \omega}{||v||^2} \iota_v F \wedge \star  \iota_v F  \nn \\
=  2 \cos^2 \frac{\omega}{2} ~F^+ \wedge \star F^+ +  2 \sin^2 \frac{\omega}{2}~ F^- \wedge \star F^- - \frac{\sin^2 \omega}{||v||^2} \iota_v F \wedge \star  \iota_v F~, \label{identity-FF-proj}
\eea
where $F^\pm = 1/2 (1\pm \star )F$. These identities will play a crucial role in the following.

In summary, on a 4D manifold with a globally defined vector field $v$ with isolated fixed points (and additional data at the fixed points as described above) and a metric,  we have an alternative decomposition of two forms $\Omega^2$ given by  the  projectors $P_\omega^+ +   P_\omega^- = 1$
\bea
\Omega^2 = P^+_\omega \Omega^2 \oplus P_\omega^- \Omega^2~.\label{decomp-forms-alt}
\eea
 This provides an alternative decomposition of $\Omega^2$ into two orthogonal subspaces $P^{\pm}_\omega \Omega^2$ with respect to the standard scalar product. Throughout the paper we will use these notations for the  decomposition of two forms \eqref{decomp-forms-alt} while keeping $\Omega^{2\pm}$ for the standard decomposition into self-dual/anti-self-dual spaces.
In what follows we will assume that $v$ is a Killing vector field so that this decomposition is preserved by ${\cal L}_v$ (provided that $\cos \omega$ is invariant along $v$).  The typical setting we have in mind is that $v$ arises from some $T^2$-action on $M$ with only isolated fixed points.

\subsection{Cohomological complex}\label{subs:cohom-complex}

Donaldson-Witten theory can be defined as a cohomological field theory \cite{Witten:1988ze, Atiyah:1990tm} which is  related to the Donaldon invariants of four manifolds.
If a four manifold admits the action of a group (for example $T^2$), one can further define an equivariant extension of Donaldson-Witten theory \cite{Nekrasov:2003rj}.
Let us review some basic facts about equivariant  Donaldson-Witten cohomological field theory.
Assuming that we have a Killing vector field $v$, we can define the following odd transformations
\bea
\label{complextransf}
&& \delta A =i \Psi~ , \nn \\
&& \delta \Psi =  \iota_{v} F  +  i d_A \phi  ~, \nn \\
&& \delta \phi =  \iota_{v} \Psi~, \nn  \\
&& \delta \varphi = i \eta~, \label{eq:cohomcomplex4d2} \\
&& \delta \eta =  L_{v}^A \varphi -  [ \phi, \varphi ] ~,\nn \\
&& \delta \chi = H~ , \nn \\
&& \delta H = i L_{v}^A \chi - i [\phi,\chi ] ~, \nn
\eea
where $A$ is a gauge connection, $\Psi$ is an odd one-form, $\phi$ and $\varphi$ are even scalars, $\eta$ is an odd scalar, $\chi$ is a self-dual odd two form and $H$ is a self-dual even two form. All these fields (except $A$) take values in the adjoint representation of the gauge group.
Throughout the paper we use the following conventions: the covariant derivative is defined as $d_A = d - i [A,~]$, the covariant version of the Lie derivative is defined as follows
$${\cal L}_v^A= d_A \iota_v + \iota_v d_A = {\cal L}_v - i [ \iota_v A, ~]$$
and the field strength as $F= dA - i A^2$.
The square of the transformations (\ref{eq:cohomcomplex4d2}) is given by
\bea
\delta^2 = i {\cal L}_v - G_{ \phi + i \iota_v A}~,\label{algebra-cohom-notfull}
\eea
where $G_{\epsilon}$ stands for a gauge transformation acting on the gauge field as
\bea
G_\epsilon A = d_A \epsilon~,
\eea
and on all other field in the adjoint as
\bea
G_\epsilon \bullet =  i [\epsilon, \bullet ]~.
\eea
We can further add the ghost $c$, anti-ghost $\bar{c}$ and Lagrangian multiplier $b$ and extend all transformations such that they square to the Lie derivative only (see appendix \ref{app2}). The cohomological theory given by the transformations (\ref{eq:cohomcomplex4d2}) is not uniquely defined since we did not specify any reality conditions for the bosonic fields $\phi$ and $\varphi$. From standard supersymmetry considerations we know that $\phi$ and $\varphi$ cannot be two independent complex fields. Suitable reality conditions cannot be fixed by cohomological considerations alone. Later on we will see that supersymmetry together with the positivity of the supersymmetric Yang-Mills action will fix the reality conditions for $\phi$ and $\varphi$.

So far we have reviewed the definition of  equivariant  Donaldson-Witten cohomological field theory.  Now we will describe its modification. Consider a four manifold with a Killing vector field with isolated fixed points, and specify any distribution of pluses/minuses over the fixed points. According to the discussion in the previous subsection we can associate to these data an orthogonal decomposition of two forms $P^+_\omega \Omega^2 \oplus P_\omega^- \Omega^2$ (\ref{decomp-forms-alt}).  Thus we can consider the transformations (\ref{eq:cohomcomplex4d2}) acting on the same set of fields except that we impose new conditions on $\chi$ and $H$
\be
P^+_\omega \chi = \chi~, \qquad    P^+_\omega H = H~.
\ee
Since ${\cal L}_v$ preserves the spaces $P^{\pm}_\omega \Omega^2$, the transformations satisfy the algebra
(\ref{algebra-cohom-notfull}) as before.
In this way we defined a new cohomological field theory which is ultimately related to supersymmetric Yang-Mills as will be explained in the next section.
If we distribute only pluses (resp.~minuses) for all fixed points, pick up a  non-trivial function $\cos \omega >0$ (resp. $\cos \omega <0$) and construct the projectors $P_\omega^{\pm}$, then one can globally rotate $P^+_\omega \Omega^2$ to $\Omega^{2+}$, bringing us back to equivariant Donaldson-Witten theory.
However, this is impossible if we have both pluses and minuses over different fixed points. Hence generically $P^+_\omega \Omega^2$ is not isomorphic to $\Omega^{2+}$ and the corresponding cohomological theory is not related to standard equivariant Donaldson-Witten theory.
It is important to remember that  the bundle $P^+_\omega \Omega^2$ is defined  up to isomorphism,
so that in general we could use different vector fields in the transformations (\ref{complextransf}) and to define the projector (\ref{projector}).
To avoid confusion we will postpone discussing this point to section \ref{sec:changingvector}.

The observables in the new theory are constructed  in the same way as in equivariant Donaldson-Witten theory, see \cite{Nekrasov:2002qd, Nekrasov:2003rj}. Let us comment on a particular class of observables that will be relevant later on.
Using  the transformations  (\ref{eq:cohomcomplex4d2}) one can observe that
\bea
\delta \Big ( \phi + \Psi + F \Big ) = (i d_A + \iota_v) \Big ( \phi + \Psi + F \Big )~.
\eea
This implies the following
\bea
\delta \Tr \Big ( \phi + \Psi + F \Big )^k = (i d + \iota_v) \Tr \Big ( \phi + \Psi + F \Big )^k~,
\eea
where $\Tr$ can be replaced by any Ad-invariant polynomial over a Lie algebra. If we pick an equivariantly closed form $\Omega$ on $M$ we can construct the observable
\bea
\int\limits_M \Omega \wedge \Tr \Big ( \phi + \Psi + F \Big )^k~,
\eea
which is annihilated by $\delta$.
This observable depends only on the class of $\Omega$ in the equivariant cohomology since the shift $\Omega \rightarrow \Omega + (id + \iota_v) [ ...]$ will lead to a $\delta$-exact shift in the observable.
Here the imaginary $i$ in front of $d$ is due to our conventions and does not play any essential role.
Once reality conditions are set we can discuss the reality and positivity of the bosonic part of these observables.  Among all observables we will be particularly  interested in
\bea
&&  \mathcal{O} =   \int\limits_M \Big ( \Omega_0 + \Omega_2 + \Omega_4 \Big ) \wedge \Tr \Big ( \phi + \Psi + F \Big )^2  \nn \\
&&  = \int\limits_M \Big ( \Tr(\phi^2) \Omega_4 + 2 \Omega_2 \wedge \Tr (\phi F) + \Omega_0 \Tr (F^2) + \Omega_2 \wedge \Tr (\Psi^2) \Big )~,\label{obs-abstract}
\eea
where $(\Omega_0 + \Omega_2 + \Omega_4 )$ is closed under $i d + \iota_v$.
As we will show in the next section, this observable is closely related to the supersymmetrized Yang-Mills action.

\section{Supersymmetry}\label{s:supersymmetry}

We start considering a 4D spin manifold $M$  equipped with a Riemannian metric $g$. We will consider the spin$^c$ case at end of section \ref{secohmvrb}. On $M$ we have left and right handed spinors $\zeta^i_\alpha$ and $\bar\chi_i^\alphadot$ transforming in the fundamental of the $SU(2)_R$ R-symmetry. Here $i$ is the $SU(2)_R$ index while $\alpha,\alphadot$ are spinor indices (conventions for spinors are spelled out in appendix \ref{app:conventions}).

In this section we construct ${\cal N}=2$ supersymmetric field theories on $M$ given the following further  data:
\begin{itemize}
\item{}{We impose that $g$ admits a smooth real Killing vector field $v$ with at most isolated fixed points $x_i$~.\footnote{ The possibility that the fixed points are not isolated is not excluded but requires a case by case analysis.}}
\item{}{A smooth function $s$ on $M$ that is positive everywhere except at a subset of the fixed points of $v$ where it vanishes. We also require that $v^\mu \partial_\mu s=0$ ($s$ is invariant along $v$) and that $\tilde s= s^{-1} ||v||^2$ is smooth.}
\end{itemize}
The theories we construct will admit one supercharge $\delta$ squaring to a translation along $v$.

Note that either $s$ or $\tilde s$ vanishes at each fixed point of $v$ because $||v||^2 = s \tilde s$.
This determines the $\pm$ sign associated to each fixed point as described in the previous section: if $\tilde s = 0$ we have a positive sign, and if $s=0$ we have a negative sign.

\subsection{Construction of global spinors}
\label{spincon}

As a first step we will use the geometrical data above to construct smooth spinors $\zeta^i_\alpha$ and $\bar\chi_i^\alphadot$  satisfying the reality conditions $({\zeta}_{i \alpha})^*= {\zeta}^{i \alpha}$ and $(\bar \chi_i^\alphadot)^*=\bar \chi^i_\alphadot $ and such that
\be
\label{bilspin}\zeta^i \zeta_i={s\over 2}~,\qquad \bar \chi^i \bar \chi_i={\tilde s\over 2}~, \qquad \bar \chi^i\bar\sigma^\mu \zeta_i ={1\over 2} v^\mu~.
\ee

We cover $M$ with open patches $U_k$, each equipped with a choice of local frame $e^a_k$ and we  assume that each fixed point of $v$  belongs to a single distinct $U_k$\footnote{This is not the same covering as in \ref{s:subs-projector} because some of the patches need not contain a fixed point of $v$.}.
In the overlap between $U_k$ and $U_l$ the frames $e^a_k$  and $e^a_l$ are related by an $SO(4)=SU(2)_l\times_{\mathbb{Z}_2} SU(2)_r$ transformation. The spinors $\zeta$ and $\chi$ are related by $SU(2)_l\times_{\mathbb{Z}_2} SU(2)_R$ and $SU(2)_r\times_{\mathbb{Z}_2} SU(2)_R$  transformations respectively. By identifying $SU(2)_l$ with $SU(2)_R$  we can construct a globally well defined topologically twisted spinor $\zeta_t$ whose component expression in each patch $U_k$ is  given by $({\zeta_t})^i_\alpha=\delta^i_\alpha$.

Away from the zeros of $s$ we can then define spinors $\zeta^i$ and ${\bar \chi}_i$ as follows:
\be
\label{defchi} \zeta^i={\sqrt{s}\over 2}\, \zeta^i_t~,\qquad  {\bar \chi}_i=  {1\over s} v^\mu \bar \sigma_\mu  \zeta_i~.
\ee
These definitions ensure that \eqref{bilspin} are satisfied and the spinors are real.
The spinor $ {\bar \chi}_i$ is singular at the fixed points where $s=0$, the bilinears \eqref{bilspin} however are everywhere smooth. In a patch $U_k$ containing a zero of $s$ the function $\tilde s$ is strictly positive and we can define smooth spinors:
\be
\label{newspin}\hat {\bar \chi}_i^\alphadot= -i {\sqrt{\tilde s}\over 2}\delta^\alphadot_i~,\qquad \hat\zeta_i= -{1\over \tilde s}v^\mu \sigma_\mu \hat{\bar \chi}_i~,
\ee
whose bilinears and reality properties are the same as those of $\bar \chi$ and $\zeta$~. It follows that the hatted spinors $\hat{\bar \chi},~\hat \zeta$ are related to $\bar \chi$ and $\zeta$ by an $SU(2)_R$ transformation
\be
\label{spinrot}\bar \chi_i= {U_i}^j \hat { \bar \chi}_j~,\qquad  \zeta_i= {U_i}^j  \hat \zeta_j~,
\ee
which is explicitly given by
\be
\label{suttr} {U_i}^j= i {v^\mu \over ||v||}{{\sigma_\mu}_i}^j~.
\ee

Let $\Sigma$ be a small three sphere surrounding a fixed point of $v$ where $s=0$. The map ${U_i}^j$ from $\Sigma$ to $SU(2)_R$ is non-singular and of degree $1$.
We can now consider  spinors that are equal to  \eqref{defchi} in all patches except those containing a zero of $s$ where the spinors are given by \eqref{newspin}. In going from a patch $U_k$ containing a zero of $s$ and a second patch $U_l$ the $SU(2)_R$ transformation is \eqref{suttr} followed by that  corresponding to topological twisting.
This construction results in spinors that are smooth everywhere on $M$ and whose bilinears are given by  \eqref{bilspin}.

\subsubsection{Spinor bilinears}
\label{secspinbil}

Using the spinors $\zeta$ and $\bar\chi$  we can form other spinor bilinears besides $v^\mu$ and $s,\tilde s$,
\be\label{bilspi}
\Theta_{\mu\nu}^{(ij)}=\zeta^i \sigma_{\mu\nu}\zeta^j~,\qquad \widetilde \Theta^{(ij)}_{\mu\nu}=\bar\chi^i \bar \sigma^{\mu\nu} \bar \chi^j~,\qquad v^{(ij)}_\mu= \zeta^i \sigma_\mu \bar\chi^j+\zeta^j \sigma_\mu \bar\chi^i.
\ee
These are forms valued in the adjoint of $SU(2)_R$.

The vector field $v$ allows us to construct the family of projectors $P_\omega^+$ in \eqref{projector}.
We will impose that the function $\omega\in [0,\pi]$ behaves as follows near the fixed points
\be\label{bdycond}
\omega=o(\sqrt{\tilde s}\,) {\rm ~~for~}\tilde s\sim 0~,\qquad \omega=\pi- o(\sqrt{s}) {\rm ~~for~} s\sim 0~.
\ee
This guarantees smoothness of $P^+_\omega$. According to the discussion in section \ref{s:subs-projector} we see that at the fixed points with $\tilde{s}=0$ the projector collapses to the self-duality projector and at the fixed points with $s=0$ to the anti-self-duality projector. One specific choice of $\omega$, which is is completely specified by the Killing spinors,  will be referred below as  ``canonical''
 \be
 \label{canchoice}\cos\omega_{\rm{c}}={s-\tilde s\over s+\tilde s}~.
 \ee
Using $\Theta$ and $\widetilde \Theta$ in \eqref{bilspi} we can form the combination
\be\label{defthetahat}\widehat\Theta^{i j}_{\mu\nu}= {4\over 1+\cos^2\!\omega} \left({\cos^2(\omega/ 2)\over s}\Theta^{i j}_{\mu\nu}+ {\sin^2(\omega/2)\over \tilde s}\widetilde\Theta^{i j}_{\mu\nu}\right)~,
\ee
which is everywhere smooth because of \eqref{bdycond} and enjoys the following properties
\be\label{thetaprop}P_\omega^+ \widehat \Theta^{i j}=\Theta^{i j}~,\qquad \widehat\Theta^{i j}_{\mu\nu}\widehat\Theta_{i j}^{\rho\lambda}={1\over 1+\cos^2\!\omega}{P^+_\omega}_{\mu\nu}^{\rho\lambda}~, \qquad \widehat \Theta^{i j}_{\mu\nu}\widehat\Theta_{k l}^{\mu\nu}={\delta^i_k\delta^j_l+\delta^i_l\delta^j_k\over 2(1+\cos^2\!\omega)}~.
\ee

\subsection{Solving the Killing spinor equations}\label{sec:killingspinorsol}

Given a  supersymmetric field theory in flat space we can couple it to a supergravity background. The conditions for the background to preserve supersymmetry are then encoded in generalised Killing spinor equations that the supersymmetry variation parameters (in our case $\zeta$ and $\bar \chi$) have to satisfy \cite{Festuccia:2011ws,Dumitrescu:2016ltq}. We will consider ${\cal N}=2$ theories with a conserved $SU(2)_R$ current whose supercurrent multiplet was studied by Sohnius \cite{Sohnius:1978pk}. These  theories couple to the  ${\cal N}=2$ Poincar\'e supergravity described in \cite{DEWIT1980186,DEWIT198177,DEWIT1985569,freedman_van_proeyen_2012}.  A supergravity background (see for instance \cite{Butter:2015tra}) is specified by the metric $g$, a choice of $SU(2)_R$ connection ${{V_\mu}^i}_j$, a scalar N, a one form $G_\mu$, a two form $W_{\mu\nu}$, a scalar $SU(2)_R$ triplet $S_{i j}$, and finally an closed two form ${\mathcal F}_{\mu\nu}$ corresponding to the graviphoton field strength.

There are two sets of Killing spinor equations. The first  arises from setting the variation of the gravitino to zero
\begin{align}\label{killingn}
&(D_\mu -i G_\mu) \zeta_{i}-{i\over 2} W^+_{\mu\rho} \sigma^{\rho}{\bar \chi}_{i} -{i\over 2} \sigma_\mu {\bar \eta}_i=0~,\cr
&(D_\mu + i G_\mu) {\bar \chi}^{i}+ {i\over 2} W^-_{\mu\rho} {\bar \sigma}^\rho{ \zeta^{i}} -{i\over 2} {\bar \sigma_\mu} {\eta}^i=0~,
\end{align}
while the second set is obtained setting the variation of the dilatino to zero
\begin{align}\label{dilatino}
&\Big(N-{1\over 6} R\Big)\bar \chi^i=4i \partial_\mu G_\nu \bar \sigma^{\mu\nu} \bar \chi^i+{i} \big(\nabla^\mu+2 i G^\mu\big) W^{-}_{\mu\nu} \bar \sigma^\nu\zeta^i +i \bar \sigma^\mu\big(D_\mu+{i} G_\mu\big) \eta^i ~,\cr
&\Big(N-{1\over 6} R\Big)\zeta_i=-4i \partial_\mu G_\nu \bar \sigma^{\mu\nu} \zeta_i-{i} \big(\nabla^\mu-2i G^\mu\big) W^{+}_{\mu\nu} \sigma^\nu\bar \chi_i +i \sigma^\mu\big(D_\mu-{i} G_\mu\big) \bar \eta_i ~.
\end{align}
In the above equations the covariant derivative $D_\mu$ includes the $SU(2)_R$ connection ${{V_\mu}^i}_j$ so that, for instance, $D_\mu \zeta_i =\nabla_\mu \zeta_i - {V_\mu^j}_i \zeta_j$.
The spinors $\eta_i$ and $\bar \eta_i$ are given by
\begin{align}
\label{defeta}
& \eta_i=({\mathcal F}^+-W^+) \zeta_i-2 G_\mu\sigma^\mu \bar \chi_i- S_{i j} \zeta^j~,\nonumber \\[2pt]
& \bar \eta^i= -({\mathcal F}^-- W^-) \bar \chi^i +2 G_\mu \bar \sigma^\mu \zeta^i-S^{i j} \bar \chi_j~.
\end{align}
We used the shorthand notation $W^+={1\over 2} W_{\mu\nu} \sigma^{\mu\nu}$ and  $W^-={1\over 2} W_{\mu\nu}\bar  \sigma^{\mu\nu}$ (similarly for $\mathcal F$).

Provided some integrability conditions (and smoothness requirements) are satisfied, the equations above can be solved (albeit non uniquely) in terms of the spinors $\zeta_i$ and $\bar \chi_i$ constructed above. The resulting supergravity background is smooth. For the case of topological twisting this was implemented in \cite{Karlhede:1988ax}.

In order to solve the first set of two equations in \eqref{killingn} we need $v$ to be a Killing vector field and that $s$ (hence $\tilde s$) is invariant along $v$. The solution reads
\begin{align}
\label{solTVaux}
W_{\mu\nu} =& {i \over s+\tilde s} (\partial_\mu v_\nu-\partial_\nu v_\mu)-{2i \over (s+\tilde s)^2} {\epsilon_{\mu\nu\rho}}^{\lambda}v^\rho \partial_\lambda(s-\tilde s)-{4 \over s+\tilde s}{\epsilon_{\mu\nu\rho}}^\lambda v^\rho G_\lambda+\cr
& +{s-\tilde s\over (s+\tilde s)^2}{\epsilon_{\mu\nu\rho}}^\lambda v^\rho b_\lambda +{1\over s+\tilde s}(v_{\mu} b_{\nu} -v_{\nu}b_{\mu})~,\cr
(V_\mu)_{i j}=&{4\over s+\tilde s}\left(\zeta_{(i}\nabla_\mu \zeta_{j)}+\bar \chi_{(i}\nabla_\mu \bar \chi_{j)}\right)+{4\over s+\tilde s} \left(2 i G_\nu-{\partial_\nu(s-\tilde s)\over (s+\tilde s)}\right){(\Theta_{i j}-\widetilde \Theta_{i j})^\nu}_\mu+\cr&+ {4i\over (s+\tilde s)^2}b_\nu {( \tilde s\, \Theta_{i j}+s\,\widetilde  \Theta_{i j} )^\nu}_\mu~,
\end{align}
where $b_\mu$ is a one form on $M$ satisfying $v^\mu b_\mu=0$.
We make the following remarks:
\begin{itemize}
\item{ $s+\tilde s>0$ everywhere, hence, because the spinors $\zeta_i, \bar\chi_i$ and their  bilinears defined in section \ref{secspinbil}  are smooth so are $W$ and $V$.}
\item{The one forms $b_\mu$ and $G_\mu$ are left undetermined and parametrise different solutions of  \eqref{killingn}.\footnote{We can describe the freedom parametrized by $b_\mu$ via a two form $U_{\mu\nu}$ that satisfies $P^{+}_{\omega_{\rm can}} U=U$ where the canonical choice for $\omega$ is as in \eqref{canchoice}. This is somewhat more general because a $b_\mu$ that is singular at the fixed points can correspond to a smooth $U$, but it makes explicit expressions more complicated.} We could use this freedom to set $W_{\mu\nu}$ to zero but generically the required $G_\mu$ and $b_\mu$ would not be smooth. }
\item{The expression for the $SU(2)_R$ background gauge field $(V_\mu)_{i j}$ in a given patch is generally complex. However, in transitioning from patch to patch, as described in section \ref{spincon},  $(V_\mu)_{i j}$ changes by a real $SU(2)_R$ gauge transformation. Hence the imaginary part of $(V_\mu)_{i j}$ is a globally defined one form in the adjoint of $SU(2)_R$.}
\item{In the large volume limit  the supergravity background fields approach their values in flat space. Indeed the background fields $W_{\mu\nu}$ and $(V_\mu)_{i j}$ scale as ${1\over r}$ where $r$ is the overall size of the manifold $M$ (assuming that $G_\mu$ and $b_\mu$ are also chosen of order~$r^{-1}$).}

\end{itemize}

Next we consider the second set of equations \eqref{dilatino}. They can all be solved provided that $G_\mu$ and $b_\mu$ are invariant along $v$ and just determine  the scalar $N$. This fact was already noted in \cite{Klare:2013dka,Pestun:2014mja}. The solution for $N$ is  everywhere smooth and of order $r^{-2}$. The explicit expression for N is rather lengthy and can be found in the appendix \ref{appE},  equation \eqref{solN}.

We can now use  \eqref{defeta} to determine the remaining supergravity background fields. A solution for the closed two form ${\mathcal F}_{\mu\nu}$ is given by
\be
\label{solF}
{\mathcal F}_{\mu\nu}=i \partial_{\mu}\Big({s+\tilde s-K\over s \tilde s} v_\nu\Big)-i\partial_{\nu}\Big({s+\tilde s-K\over s \tilde s} v_\mu\Big)~,
\ee
where $K$ is a constant. Here ${\mathcal F}_{\mu\nu}$ is not just closed but exact. In order to guarantee that  ${\mathcal F}_{\mu\nu}$ is smooth we need to impose the extra condition that $s+\tilde s$ approaches the same constant $K$ at all fixed points fast enough. This can always be arranged. In principle we could ask that $s+\tilde s=K$ everywhere on $M$ and not just at the fixed points. This normalization, however, imposes some restriction on the metric. Namely if $s$ is zero at some but not all of the fixed points of $v$, then  $s=\tilde s$ on some codimension 1 locus on $M$ and the norm of the Killing vector $v$ would have to attain its extremum there. This can always be arranged in a given conformal class. In the following we will not assume that $s+\tilde s$ is constant.

The solution \eqref{solF} is not the most general. We could add to it any closed two form $\hat f$ satisfying $\iota_v \hat f=0$. In specific cases it may be that this freedom allows to relax the condition that $s+\tilde s=K$ at the fixed points of $v$.

The scalar triplet $S_{i j}$ is determined as well. The resulting expression is presented in appendix \ref{appE} , equation \eqref{solS}. Smoothness of $S_{i j}$ also requires $s+\tilde s$ to approach $K$ at fixed points fast enough.

\subsection{Vector multiplet}

An ${\cal N}=2$ vector multiplet comprises, in addition to the gauge field $A_\mu$, of a complex scalar $X$, gauginos $\lambda_{i\alpha}, \tilde \lambda^i_{\dot\alpha}$ transforming in the fundamental of $SU(2)_R$ and an auxiliary real scalar $SU(2)_R$ triplet $D_{i j}$. All these fields (except $A$) transform in the adjoint of the gauge group.
This multiplet can be coupled to background ${\cal N}=2$ supergravity as described in \cite{Butter:2015tra} (see also \cite{DEWIT1980186,DEWIT198177,DEWIT1985569,freedman_van_proeyen_2012}).

\subsubsection{Supersymmetry algebra}

In a supersymmetric  background the supersymmetry variations are given by
\begin{eqnarray}
\label{vmulvr}
&& \delta A_{\mu}=i\zeta_i \sigma_\mu {\bar \lambda}^i+i {\bar \chi}^i {\bar \sigma}_\mu \lambda_i~, \nonumber \\
&&\delta {\bar X}= {\bar \chi}^i{\bar \lambda_i}~,\qquad \delta { X}= -{ \zeta}_i{\lambda^i} ~, \nonumber \\
&&\delta D_{i j}= i \zeta_i \sigma^\mu \big( D_\mu\! +i  G_\mu \big){\bar \lambda}_j -i {\bar \chi}_i \bar \sigma^{\mu}\big( D_\mu\! -i G_\mu \big) \lambda_j+2i  [X,\bar \chi_i \bar \lambda_j]+2i  [\bar X, \zeta_i \lambda_j] + (i \leftrightarrow j)~, \nonumber\\
&&\delta {\bar \lambda}^{i}=2 i (D_\mu +2i G_\mu){\bar X} {\bar \sigma^\mu\zeta^i}\!+\!2 \big(F^-\!- X\, W^-\big)\bar \chi^i-D^{i j}\bar \chi_j-2i  [X, \bar X] \bar \chi^i+2 \bar X \bar \eta^i~, \nonumber \\
&& \delta \lambda_{i}=-2 i (D_\mu-2iG_\mu)  X {\sigma^\mu{\bar \chi}_i}\!+2\big(F^+\!-{\bar X}\, W^+\big) \zeta_i+D_{i j}\zeta^j+2 i  [X,\bar X] \zeta_i-2 X\eta_i~.
\end{eqnarray}
Here $F_{\mu\nu}$ is the field strength for the gauge field $A_\mu$ and we used the shorthand notation $F^+={1\over 2 } F_{\mu\nu} \sigma^{\mu\nu}$ and  $F^-={1\over 2 } F_{\mu\nu} \bar \sigma^{\mu\nu}$.

The composition of two transformations above on a field $\Phi$ in the vector multiplet results in a translation along the Killing vector field $v$ together with an $SU(2)_R$ transformation and a gauge transformation
\be
\label{salg}
\delta^2 \Phi=   i {\cal L}_{v} \Phi+ i v^\mu V_\mu \circ \Phi+  v^\mu [ A_\mu,\Phi] +i \Lambda^{\!(R)} \!\!\circ \Phi-i [s \bar X+\tilde s X,\Phi]
\ee
here ${\cal L}_{v}$ is the Lie derivative along $v$, and $\circ$ denotes that $\Phi$ is acted upon according to which  $SU(2)_R$ representation it belongs. $\Lambda^{\!(R)}$ is a $SU(2)_R$ transformation parameter:
\be
\Lambda^{\!(R)}_{i j}=\bar \chi_i \bar \sigma^\mu (D_\mu -i G_\mu)\zeta_j-\zeta_i \sigma^\mu (D_\mu+iG_\mu) \bar \chi_j+ (i\leftrightarrow j)~.
\ee

\subsubsection{Lagrangian}

From the coupling to supergravity we can also read the form of supersymmetric Lagrangians,
\begin{eqnarray}
\label{lagvec}
&{\cal L}={1\over g^2}{\rm Tr}\left[-4(D^\mu\!+2 i G^\mu){\bar X}\, (D_\mu\!-2 i G_\mu)X -{1\over 2} F_{\mu\nu} F^{\mu\nu} -i{\theta g^2 \over 32 \pi^2}\epsilon^{\mu\nu\rho\lambda}F_{\mu\nu}F_{\rho_\lambda} +{1\over 2}  D^{ i j} D_{i j} + \right.\nonumber \\[3pt]
&-4 [X,\bar X]^2+
2 F^{ \mu\nu}(X W^-_{\mu\nu} +{\bar X}W^+_{\mu\nu})-X^2 W^-_{\mu\nu}W^{-\mu\nu} \!-{\bar X}^2 W^+_{\mu\nu}W^{+\mu\nu}+ \left.4\Big({R\over 6 }  -N\Big) X{\bar X}\right] +\nonumber \\[3pt]
&-{1\over g^2}{\rm Tr}\left[ i \lambda_i  \sigma^\mu \Big(\!D_\mu\!+ i G_\mu\Big){\bar \lambda}^{i}+i \bar \lambda^i  \bar\sigma^\mu \Big(\!D_\mu\!-i G_\mu\Big){ \lambda}_{i}+2i \lambda^i [\bar X, \lambda_i]+2i \bar \lambda^i [ X, \bar \lambda_i]\right]~.
\end{eqnarray}
At short distances this Lagrangian is a small deformation of a Lagrangian for the vector multiplet in flat space. Indeed the coupling to the supergravity background introduces terms scaling as $r^{-1}$ or $r^{-2}$ with the overall size $r$ of the manifold.

In general the Lagrangian (\ref{lagvec}) may not have positive real part and using it to define the path integral is problematic.
This issue  depends on the concrete values of the background fields and it should be addressed case by case. For example for the theory on a round $S^4$ studied by \cite{Pestun:2007rz} the Lagrangian (\ref{lagvec}) is real and positive. For the theories on squashed $S^4$ considered in \cite{Hama:2012bg, Pestun:2014mja} the real part of (\ref{lagvec}) continues to be positive at least for small squashing.
 For more general backgrounds this problem becomes more complicated to analyze and it may force us to reconsider the reality conditions for some fields (or just for some modes of the fields). Another option can be that of adding appropriate $\delta$-exact terms to the Lagrangian. We leave this issue aside for now.

\subsubsection{Cohomological variables \label{secohmvrb}}

The action of supersymmetry on the components of the vector multiplet \eqref{vmulvr} and the structure of the supersymmetric Lagrangian \eqref{lagvec} are more transparent if we rewrite them using appropriate cohomological variables.
This is also how we connect the supersymmetric theory to the cohomological complex described in section \ref{subs:cohom-complex}.
The change of variables  will make use of the projector $P^{+}_\omega$ defined in \eqref{projector}.
Here we select the canonical choice for the function $\omega$ defined in \eqref{canchoice} because it results in simpler expressions.
We define the following cohomological fields:
\be
\begin{aligned}
\label{twisted}
&\eta=\zeta_i \lambda^i+\bar \chi^i \bar \lambda_i~,\cr
&\varphi= -i(X-\bar X)~,\cr
&\Psi_\mu= \zeta_i \sigma_\mu \bar \lambda^i+\bar \chi^i\bar \sigma_\mu \lambda_i~,\cr
&\phi= \tilde s X+ s \bar X~,\cr
&\chi_{\mu\nu}=2{s+\tilde s\over s^2+\tilde s^2}\left({ \bar \chi^i\bar\sigma_{\mu\nu}\bar \lambda_i} -{ \zeta_i \sigma_{\mu\nu} \lambda^i} +{1\over s + \tilde s} (v_{\mu} \Psi_\nu-v_\nu \Psi_\mu)\right)~,\cr
&H_{\mu\nu} = {(P^+_{\omega_c})}_{\mu\nu}^{\rho \lambda}\left[  \hat \Theta^{i j}_{\rho\lambda} D_{i j}- F_{\rho\lambda}+i{X+\bar X\over s+\tilde s} (\partial_\rho v_\lambda-\partial_\lambda v_\rho)+ \right. \cr
&\qquad \left.-{2i\over s+\tilde s} {\epsilon_{\rho\lambda\gamma}}^{\delta}v^\gamma\left(\Big(D_\delta-2iG_\delta -i{\tilde s\over s+\tilde s}\, b_\delta \Big)X-\Big(D_\delta+2iG_\delta -i {s\over s+\tilde s}\, b_\delta\Big)\bar X\right) \right]~.
\end{aligned}
\ee
We note the following
\begin{itemize}
\item{With the standard reality conditions on the scalar fields $\bar X= X^\dagger$ the field $\varphi$ is real while the field $\phi$ is complex. We can write
$$
\phi= {1\over 2} (s+\tilde s) (X+\bar X)-{i\over 2}(s-\tilde s) \varphi~,
$$
hence the imaginary part of $\phi$ is determined by $\varphi$. It follows that $\phi$ and $\varphi$ together have the same degrees of freedom as the complex scalar field $X$.
}
\item{All the cohomological fields in \eqref{twisted} are differential
 forms with values in the adjoint of the gauge group. They are all singlets under the $SU(2)_R$ symmetry.}
\item{The two form $\chi_{\mu\nu}$ satisfies $(P^+_{\omega_c})\chi=\chi$ as does the two form $H_{\mu\nu}$.}
\item{$H_{\mu\nu}$ is the only twisted variable whose definition involves the auxiliary fields in the gravity multiplet and derivatives of the Killing spinor bilinears.}
\end{itemize}

The change of variables \eqref{twisted} can be inverted,
\be
\label{inversetw}
\begin{aligned}
&X={1 \over s+\tilde s}\,(\phi +i\, s \, \varphi) ~,\qquad \bar X={1 \over s+\tilde s}\,(\phi - i\, \tilde s \,\varphi)~,\\
& \bar \lambda_i ={1\over s+\tilde s}\,\Big(2 \bar \chi^j  (\Theta^{\mu\nu}_{j i}+\widetilde\Theta^{\mu\nu}_{j i} )\chi_{\mu\nu}+ \bar \sigma^\mu \zeta_i\Psi_\mu+{\bar \chi_i \eta}\Big)~,\\
&\lambda_i ={1\over s+\tilde s} \,\Big(2 \zeta^j  (\Theta^{\mu\nu}_{j i}+\widetilde\Theta^{\mu\nu}_{j i} )\chi_{\mu\nu} -\sigma^\mu \bar \chi_i\Psi_\mu-{\zeta_i \eta}\Big)~,\\
&D_{i j}=4{s^2+\tilde s^2\over (s+\tilde s)^2} \hat \Theta^{\mu\nu}_{i j}(H_{\mu\nu}-\ldots)~,
\end{aligned}
\ee
where in the last formula we subtract from $H_{\mu\nu}$ all the terms in its definition \eqref{twisted} that are not proportional to $D_{ij}$.

Supersymmetry  (\ref{vmulvr}) induces the following transformations on the cohomological fields
\be
\label{complexsusy}
\begin{aligned}
	&\delta A = i \Psi ~, \\
	&\delta \varphi = i \eta ~, \\
	&\delta \chi = H ~, \\
	&\delta \phi = \iota_v \Psi ~.
\end{aligned} \qquad
\begin{aligned}
	&\delta \Psi = \iota_v F + i d_A\phi ~, \\
	&\delta \eta = \mathcal{L}_v^A \varphi - [\phi,\varphi]~, \\
	&\delta H = i\mathcal{L}_v^A \chi - i [\phi,\chi] ~, \\
	& \
\end{aligned}
\ee	
which exactly coincide with the cohomological  transformations in \eqref{complextransf} defined in section \ref{subs:cohom-complex}.
Here ${\cal L}^A_v$ is the gauge covariant Lie derivative along $v$. Acting with $\delta$ twice reproduces the algebra \eqref{salg}.

At the beginning of this section we required the manifold $M$ to be spin. However, after switching to cohomological variables the transformations (\ref{complexsusy}) can be defined for non-spin manifolds and a choice of rank 3 bundle $P_\omega^+ \Omega^2$ (e.g., for $\mathbb{CP}^2$ with the different allocations of pluses and minuses to 3 fixed points and correspondently different $P_\omega^+ \Omega^2$).  Thus in what follows we will use the cohomological variables for the vector multiplet and consider also non-spin examples.

 Here we do not discuss hypermultiplets which can also be coupled to the supergravity background.  In a way similar to the case of topological twisting, (see e.g \cite{Labastida:2005zz} for a list of references) suitable cohomological variables can be defined also for hypermultiplets.  Since the twisted hypermultiplet will contain spinors it will be defined on spin manifolds. Coupling a $U(1)$ global symmetry acting only on the hypermultiplets to a suitable background gauge field allows to extend the twisted theory to the spin$^c$ case \cite{Marino:1996sd}. We leave the hypermultiplets to further detailed study.

\subsubsection{Lagrangian in cohomological variables}

Here we rewrite the Lagrangian \eqref{lagvec} using cohomological variables.
 For simplicity we will present the result for the canonical choice of projector resulting in the cohomological
  fields \eqref{twisted}. Later we will  comment on what would change with a different choice of projector.

The starting point is the following rewriting of the Yang-Mills Lagrangian
\be
\label{ymtws}\nonumber
{\rm Tr} \left[ F\wedge \star F 
\right]= {\rm Tr} \left[ ({1+\cos^2\!\omega})(P^{+}_{\omega} F)\wedge \star  F +{\sin^2\!\omega\over ||v||^2 } \iota_v F\wedge\star \iota_v F-\cos\omega \; F\wedge F \right]~.
\ee
Modulo $\delta$-exact contributions and the $\theta$ term, the Lagrangian \eqref{lagvec} arises from the supersymmetrization $\cal O$ of the last term in the equation above,
\be
\label{twilag}
{\cal L}~ \mathrm{Vol}_{M}  = {1\over g^2}{\cal O} - {i \theta \over 8 \pi^2}F\wedge F  +\delta\{\ldots\}~.
\ee
Specializing to the canonical choice for $\omega$ we have
\be
\begin{aligned}
	\label{obsdef}
	{\cal O}=&{\rm Tr}\left[{s-\tilde s\over s+\tilde s} F\wedge F- (  \Psi\wedge\Psi + 2 \phi F )  \wedge \left (2i \frac{s-\tilde s}{(s+\tilde s)^3} d\kappa + \frac{4i}{(s+\tilde s)^3}\kappa\wedge d(s-\tilde s) \right)
	+\right.\\
	&\left. - \phi^2 \left ( 6\frac{s-\tilde s}{(s+\tilde s)^5} d\kappa\wedge d\kappa + \frac{24}{(s+\tilde s)^5} \kappa\wedge d\kappa \wedge d(s-\tilde s) \right )\right]~,
\end{aligned}
\ee
where $\kappa = g(v)$.

We can identify $\cal O$ with the equivariant observable \eqref{obs-abstract}.
We write $h = \cos \go_c = \frac{s-\tilde s}{s+\tilde s}$ and determine the multi-form $\Omega = \Omega_0 + \Omega_2 +\Omega_4$ to be\be\label{eq:equivform1}
\begin{aligned}
&	\Omega_0 = h~ , \\
&	\Omega_2 =- 2i \frac{s-\tilde s}{(s+\tilde s)^3} d\kappa - \frac{4i}{(s+\tilde s)^3}\kappa\wedge d(s-\tilde s) ~, \\
&	\Omega_4 =-  6\frac{s-\tilde s}{(s+\tilde s)^5} d\kappa\wedge d\kappa - \frac{24}{(s+\tilde s)^5} \kappa\wedge d\kappa \wedge d(s-\tilde s) ~.
\end{aligned}
\ee
One can check that this is equivariantly closed, $(id+\iota_v)(\Omega_0+\Omega_2+\Omega_4) = 0$.
All the  forms $\Omega_i$ in \eqref{eq:equivform1} are  everywhere non-singular.

\subsubsection{\label{exsf}Example: $S^4$}

As an example we consider here the equivariantly closed form \eqref{obsdef} on the round $S^4$.
For the coordinates, metric and other conventions see appendix \ref{app:S4}.
We choose  $v= \partial_\alpha + \partial_\beta$ and the functions $s,\tilde s$ and $h$ to be
\be
	s=2\cos^2\frac{\theta}{2}~, \qquad \tilde s= 2 \sin^2 \frac{\theta}{2}~, \qquad h=\frac{s-\tilde s}{s+\tilde s} = \cos\theta~ ,
\ee
hence $h$ goes to $\pm 1$ at the poles. We also note that $s+\tilde s =2$, which leads to various simplifications.
For these choices $\kappa=g(v)$ satisfies the property (\ref{roundS4-dk}) which allows us to set all the auxiliary fields in the supergravity background to zero except for the scalar $S_{ij}$. This supergravity background has eight independent Killing spinors. The corresponding supersymmetric field theory on a round $S^4$ is that studied in \cite{Pestun:2007rz}.
The equivariantly closed form \eqref{eq:equivform1} that appears in the observable \eqref{obsdef}  takes the form
\be\label{2-formS4}
\begin{split}
	\Omega =& h - {i\over 2} (h d\kappa+ 2\kappa\wedge dh) - (\frac{3}{8} h d\kappa \wedge d\kappa + \frac 3 2 \kappa\wedge d\kappa\wedge dh   )   \\
=&	\cos \theta - i (   \sin\theta \; d\theta \wedge (x d\alpha + (x-1)  d\beta )
	+ \frac i 2 \cos\theta\sin^2\theta \; dx\wedge( d\alpha +d\beta )) \\
	&+  \frac{3}{2} \sin^3 \theta \; d\theta\wedge dx \wedge d\alpha\wedge d\beta~.
\end{split}
\ee
In particular we note that its top component is equal to $3\,{\rm Vol}_{S^4}$.  Thus $\Omega$ is an equivariant extension of the canonical volume form.
Indeed one can argue that the  lowest component $\Omega_0$ of the equivariant extension $\Omega$ of any volume form on $S^4$ (e.g., that corresponding to the squashed $S^4$ considered in \cite{Hama:2012bg, Pestun:2014mja} ) always has a different sign at the two poles.

 Applying our observations to Pestun's theory on a round $S^4$ \cite{Pestun:2007rz}, we can see that modulo the $\theta$-term the supersymmetric action can be  rewritten as
 \bea
  S_{S^4} = \int\limits_{S^4} {\cal L}~ {\rm Vol}_{S^4} = \frac{1}{g^2} \int\limits_{S^4} (\cos \theta + \Omega_2 + 3 {\rm Vol}_{S^4}) \Tr (\phi + \Psi + F)^2 + \delta\{\ldots\}~,
 \eea
where the two form $\Omega_2$ is defined in (\ref{2-formS4}). If we just  consider  the gauge sector and remember that ${\cal L}$ is positive, we see that, due to $\cos \theta$, supersymmetric configurations with positive second Chern number will be favoured around the north pole and those with negative second Chern number around the south pole.  This is just an heuristic argument that indicates the flipping behaviour realized in \cite{Pestun:2007rz} on purely cohomological grounds.

\section{Deformations}\label{secdefs}

 In defining the theory in sections \ref{s:coh-complex} and \ref{s:supersymmetry} we have used various geometrical and non-geometrical data (e.g., the values of background fields). In this section we determine what part of these data the partition function of the theory depends on. We will use the language of cohomological field theory.

When we define the fields in  cohomological field theory we use the decomposition of two-forms into orthogonal bundles $P^+_\omega \Omega^2 \oplus P_\omega^- \Omega^2$. The fields in the complex are acted upon by $\delta_v$ in (\ref{eq:cohomcomplex4d2})  which  depends on the vector field $v$. This action preserves the decomposition above.

 As we have discussed in section \ref{s:subs-projector}  the bundle $P^+_\omega \Omega^2$ is defined up to isomorphisms by a given distribution of pluses and minuses over the fixed points of $v$.  Thus if $P^+_\omega \Omega^2$ and $\tilde P^+_\omega \Omega^2$ are isomorphic and ${\cal L}_v$ preserves this isomorphism we can redefine the fields $\chi$ and $H$ in $P^+_\omega \Omega^2$ to fields  $\tilde \chi$ and $\tilde H$ in ${\tilde P}^+_\omega \Omega^2$ (very much in analogy with the logic for standard Donaldson-Witten theory \cite{Witten:1988ze}).  This redefinition does not affect the partition function since $\chi$ and $H$ enter the Lagrangian only through $\delta_v$-exact terms.

Various data enters the Lagrangian (\ref{twilag}) either through the cohomological observable ${\cal O}$ or through $\delta_v$-exact terms.  The observable ${\cal O}$ depends only on the equivariant cohomology class and not on its specific representative. Finally varying any data  that enters only in $\delta_v$-exact terms without changing the action of $\delta_v$ does not modify the value of the path integral.

Below we will comment in more detail on various deformations of the theory of the different kinds described above. We will also consider separately deformations of $\delta_v$.

\subsection{Varying $h=\cos\go$}

Changing the function $\go$ (or equivalently changing the function $h=\cos\go$)  has two effects.
First since $\cos \go$ enters the projector $P_\omega^+$ we need to redefine the cohomological fields $\chi$ and $H$ in \eqref{twisted}.
Secondly the function $h=\cos\omega$ also enters explicitly in various terms in the Lagrangian. We will show that the partition function depends only on the values of $h$ at the fixed points.

 The subbundle of two forms $P^+_\omega \Omega^2$ is determined up to isomorphism once we fix the values of $\cos\go$ at the fixed points. When changing $\cos\go$ away from the fixed points $\chi$ and $H$ should be modified appropriately as described above. The transformations can be found using the formalism developed in section \ref{s:subs-projector}. We will not have use for their explicit expressions, which are not illuminating. If the variation of $\cos\go$ is invariant along $v$ the modification of $\chi$ and $H$ commutes with $\delta_v$. Hence these variations will not affect the value of the partition function.

In the Lagrangian (\ref{twilag}) $h= \cos\omega$ enters through the cohomological observable $\cal O$ which depends on the equivariantly closed
form $\Omega$. Varying  $h$ will determine a corresponding change $\Delta \Omega$ which is also equivariantly closed. Away from the fixed points $\Delta \Omega$ is equivariantly exact.
Indeed we can write
 \be
  \Delta \Omega = (id + \iota_v) \frac{\kappa \wedge \Delta \Omega}{(id + \iota_v)\kappa}~.
 \ee
This expression is everywhere non-singular provided that the variation of $h$ and hence $\Delta \Omega$ vanishes fast enough at the fixed points. Under this condition, making use of the property
  \bea
    \int (id +\iota_v) (...) \Tr (\phi + \Psi + F)^2 = \delta \{ ... \}~,
  \eea
we find that the variation of the cohomological observable $\cal O$ determined by $\Delta \Omega$ is $\delta$-exact. Hence the partition function does not depend on the value of $h$ away from the fixed points. It follows that in principle we can set $h$ to zero everywhere except at the fixed points, this however, is a highly singular representative of the equivariant cohomology class.

\subsection{Varying the metric and other supergravity fields}

We can consider changing the metric on the four manifold $M$ keeping the vector field $v$ fixed. The variation of the metric needs to satisfy the constraint that the vector field $v$ stays Killing. As we change the metric we can keep fixed all the cohmological fields except for $\chi$ and $H$. These need to be varied so that they continue to be in the kernel of $P^-_\omega $ with the new metric. Because $v$ is Killing the variations of   $\chi$ and $H$ commute with $\delta$ hence they will not affect the partition function. The metric also enters explicitly in (\ref{twilag}) through $\delta$-exact terms. Therefore we can conclude that the path integral is independent of the metric provided that $\chi$ and $H$ are changed appropriately.

A similar reasoning can be followed to analyze the freedom in choosing the supergravity background fields. This is parametrized by $G_\mu$ and $b_\mu$ introduced in \eqref{solTVaux}. As we change the supergravity background all the cohmological fields in \eqref{twisted} are unchanged except for $H$. As long as $G_\mu$ and $b_\mu$ are kept invariant along $v$ the change in $H$ commutes with $\delta$. Because $H$ only enters the action through $\delta$-exact terms, the partition function is independent on the freedom in the supergravity background fields.

\subsection{Other choices of projector}\label{sec:changingvector}

Supersymmetry squares to a translation along the Killing vector field $v$. The projectors $P^{\pm}_\omega$ also depend on a choice of vector field, which we have implicitly assumed to be $v$ when defining the cohomological fields in \eqref{twisted}.
In principle the vector field used in defining the projector need not be the same as $v$. However, the projectors $P^{\pm}_\omega$ have to be invariant along $v$ to ensure compatibility with supersymmetry.

In many cases of interest the Killing vector field $v$ is a real linear combination of  two commuting Killing vector fields $v_1$ and $v_2$
\be
\label{relvd}
v= \epsilon_1 v_1+ \epsilon_2 v_2~\qquad \epsilon_1,\epsilon_2 \in \mathbb{R}~.
\ee
The two Killing vector fields $v_1$ and $v_2$ generate a torus action on ${M}$ with isolated fixed points. For generic values of $\epsilon_1$ and $\epsilon_2$ the orbits of $v$ are not closed and fill a torus in ${M}$.

We can define new projectors $\tilde P^{\pm}_\omega$ using a linear combination $\tilde v$ of $v_1$ and $v_2$ that is different from $v$. We use the same function $\cos\omega$ to define both sets of projectors hence the subbundles of two forms $\tilde P^+_\go \Omega^2$ and $P^+_\go \Omega^2$ are isomorphic. It follows that there is an invertible map from the cohomological fields $\chi$ and $H$ in $P^+_\go \Omega^2$ to new fields  $\tilde \chi$ and $\tilde H$ in $\tilde P^+_\go \Omega^2$. This map can be taken to commute with the action of supersymmetry because $v$ and $\tilde v$ commute. Using the same logic as in the previous examples this modification of the complex does not change the partition function.

\subsection{Complex $v$}

Here we consider again the case where $M$ admits a torus action generated by two commuting Killing vector fields $v_1$ and $v_2$. Following the procedure described  in section \ref{s:supersymmetry}, given $v$ as in \eqref{relvd} we can construct a supersymmetric field theory on $M$. We want to generalize this construction to $v$'s that are complex linear combinations  of $v_1$ and $v_2$. In order to construct the corresponding Killing spinors on $M$ we cannot directly repeat the construction of section \ref{spincon} for a number of reasons. Firstly the reality conditions on the spinors have to be relaxed.
Secondly using \eqref{suttr} for a complex $v$ would lead to complexified $SU(2)_R$ transformations in going from patch to patch.

We can instead proceed as follows. First let's consider a real $v$ as in \eqref{relvd} and choose scalar functions $s,\tilde s$ as in section \ref{spincon}.
We can then construct real spinors $({\zeta}_{i \alpha})^*= {\zeta}^{i \alpha}$ and $(\bar \chi_i^\alphadot)^*=\bar \chi^i_\alphadot $ such that
\be
\label{bilspin2}\zeta^i \zeta_i={s\over 2}~,\qquad \bar \chi^i \bar \chi_i={\tilde s\over 2}~, \qquad \bar \chi^i\bar\sigma^\mu \zeta_i ={1\over 2} v^\mu~.
\ee
Consider now a complex linear combination of $v_1$~and~$v_2$
\be
v'= \epsilon'_1 v_1+ \epsilon'_2 v_2~\qquad \epsilon'_1,\epsilon'_2\in \mathbb{C}~.
\ee
We can define the combinations
\be
\begin{aligned}
&\zeta'_i= a\, \zeta_i+ b\, v'_\mu \sigma^\mu \bar \chi_i~,\qquad  s'= \zeta'^i \zeta'_i~,\cr
& \bar \chi'_i= c\, \bar \chi_i+ d\, v'_\mu \bar \sigma^\mu \zeta_i~,\qquad \tilde s'= \bar \chi'^i \bar \chi'_i~.
\end{aligned}
\ee
where $a,b,c,d$ are complex smooth scalar functions on $ M$. For $v'$ in an open neighbourhood of $v$ it is possible to choose these functions in such a way that
\begin{itemize}
\item{ The functions  $a, b, c, d$ are invariant along $v_1$ and $v_2$ and the combination $s'+\tilde s' $ is nowhere zero on $M$. }
\item{ The complex Killing vector field $v'$ is given by $v'^\mu= \bar \chi'^i\bar\sigma^\mu \zeta'_i $~. }
\item{ $a = 1 $ near a fixed point of the torus action where $\bar \chi_i=0$. Similarly $c =1 $ at those fixed points where $\zeta_i=0$.   This ensures that the spinors $\zeta'$ and $\zeta$ vanish at the same subset of fixed points of the torus action (and similarly for $ \bar \chi'_i$ and $\bar \chi_i$).}
\item{ $a,b,c,d$ and therefore $\zeta_i'$ and $\bar \chi_i'$ depend holomorphically on $\epsilon_1'$~and~$\epsilon_2'$~.}
\end{itemize}
Because we constructed the spinors $\zeta'_i $ and $\bar \chi'_i$ via linear combinations of $\zeta_i $ and $\bar \chi_i$, the transitions in going from patch to patch are the same as described in section \ref{spincon} for  $\zeta_i $ and $\bar \chi_i$. In particular these are real $SU(2)_R$ transformations.
We can then determine the supergravity background for which $\zeta'$ and $\bar\chi'$ are Killing spinors using formulas \eqref{solTVaux},\eqref{solF},\eqref{solS} and \eqref{solN} whose derivation did not make use of the reality conditions on the spinors.
The resulting background will depend holomorphically on $\epsilon_i'$. It is worth noting that we can add $\delta$ exact terms which are not holomorphic in the $\epsilon_i'$ to the action without changing the value of the partition function. Such terms may be useful to ensure positivity of the real part of the action.

We can also implement the deformation to complex $v$ at the level of the cohomological theory. As discussed in \ref{sec:changingvector} the vector field entering the definition of $P^{\pm}_\omega$ need not be the same as $v$. Hence when changing $v$ we can keep fixed the projectors $P^{\pm}_\omega$ used to define the subbundle of two forms that $\chi$ and $H$ belong to. The dependence on $v$ of the observable $\cal O$ in \eqref{obsdef} and of the $\delta$-variations of the cohomological variables  \eqref{complextransf}  can be taken to be holomorphic in the $\epsilon_i'$. This leads to the same conclusions as above.
In summary the partition function $Z_{M_{\epsilon_1, \epsilon_2}}$ is holomorphic with respect to $\epsilon_1$ and $\epsilon_2$ (we may refer to these as generalized squashing parameters
\footnote{This terminology is somewhat misleading. The metric and $\epsilon_{1,2}$ can be varied independently provided that $v$ is Killing. In particular one can consider a $v$ corresponding to generic $\epsilon_1$ and $\epsilon_2$ on the round $S^4$~.}).

\section{Localization calculation}\label{s:localization}
 We refer to \cite{Pestun:2016zxk} for a review of the general setup of the localization procedure. The main ingredient is that one adds to the action a positive $\gd$-exact term $t\,\gd(\cdots)$. Since this change does not affect the partition function one can send $t\to\infty$. Then the path integral reduces to finding the field configurations where $\gd(\cdots)\big|_{\textrm{bos}}=0$ and performing a Gaussian integral around them. We call these configurations the localization locus, to be analyzed in section \ref{sec_Ll}.
The Gaussian integral gives the Pfaffian of the fermion quadratic term divided by the square root of the boson quadratic term
\footnote{If one is careless with signs, one can write the Pfaffian as the square root of a determinant, and simply say that the Gaussian integral gives the square root of a super-determinant between fermions and bosons. We will use such language in what follows.}.
 ${\cal N}=2$ supersymmetry implies that there are further cancellations between the two factors: for Donaldson-Witten's twisted ${\cal N}=2$ theory, these leave only $\pm 1$. More generally, e.g. for equivariant Donaldson-Witten theory, regardless of the details of the boson/fermion quadratic terms, one gets after cancellation the superdeterminant of (using \eqref{delta_2_full})
\bea
\gd^2\big|_{\textrm{Localization locus}}= i {\cal L}_v + G_{a_0} ~,
\eea
taken over half of the fields in the cohomology complex. More precisely $\gd$ has the structure $\gd q=p$, $\gd p=(i {\cal L}_v + G_{a_0})q$, and the superdeterminant is over $q$'s only. In our case the relevant cohomological transformations are presented in Appendix \ref{app2} and the
 $q$'s are $(A, \varphi, \chi, c, \bar{c}, \bar{a}_0, b_0)$.

We will analyze the localization locus next, then explain the computation of the super-determinant in section \ref{sec_1loop}. Finally we will comment on the structure of the full answer in section \ref{ss:full-answer}. For the rest of this section we take the gauge group to be $SU(N)$.

\subsection{Localization locus}\label{sec_Ll}

In order to localize the theory we add to the Lagrangian supersymmetric $\delta$-exact terms. It is important that on the integration contour these terms are positive semidefinite. We regard all fields in the theory as complex valued and specify the integration contour as a half dimensional subspace of the space of complexified fields.
We choose this contour to be compatible with the natural reality conditions for the dynamical bosonic fields in the Lagrangian \eqref{lagvec}. Namely $F$ is Hermitean and
\begin{align}
\label{contour}
\varphi \in \mathbb{R}~,\quad \tilde \phi= \phi +{i\over 2}(s-\tilde s)\varphi \in  \mathbb{R}~.
\end{align}
The two form $H$ is an auxiliary field and its integration contour is chosen so that the corresponding gaussian integral converges.

The localization terms we consider are
\begin{align}
{\cal L}_{\rm loc}\mathrm{Vol}_{M}=\delta\, {\rm Tr}\Big[2 (2\Omega - H )\wedge\star\chi + (\iota_v F - i d_A(\phi + i(s-\tilde s) \varphi ) )\wedge\star\Psi + ( \iota_v d_A\varphi + [\phi,\varphi]) \eta  \Big] .\nonumber
\end{align}
Here $\Omega(\Phi)$ is a two form valued in the adjoint of the gauge group that depends on the bosonic fields in the theory $\Phi$. For the localization terms above to be supersymmetric we need that ${\cal L}_v \Omega(\Phi)= \Omega ({\cal L}_v \Phi)$.  Moreover we will require that $\Omega$ is real on the integration contour.

\noindent The bosonic terms in ${\cal L}_{\rm loc}$ are
 \be
 \begin{aligned}
 \label{locbos}
 {\cal L}^B_{\rm loc} \mathrm{Vol}_{M}&={\rm Tr}\left[ {1\over 4}\big(2 \iota_v F+d_A((s-\tilde s)\varphi)\big)\wedge \star\big(2 \iota_v F + d_A((s-\tilde s)\varphi)\big)+ d_A \tilde \phi \wedge \star d_A\tilde \phi+\right. \\
  &+\left. \mathrm{Vol}_{M}\left(\big(\iota_v d_A \varphi\big)^2- [\tilde\phi,\varphi]^2\right) +2 P^+_{\go} \Omega\wedge\star \Omega- 2 P^+_{\go} \big(H - \Omega\big)\wedge \star \big(H- \Omega\big)\right]~.
 \end{aligned}
 \ee
The last term is set to zero by integrating over the auxiliary field $H$. The remaining terms are positive definite on the integration contour \eqref{contour} so that the path integral  localizes to field configurations such that
\begin{align}
\label{locus}
&[\tilde\phi,\varphi]=0~,\quad \iota_v d_A \varphi=0~,\quad   d_A \tilde \phi=0~,\\
\label{locusa}
& 2 \iota_v F +d_A((s-{\tilde s})\varphi)=0~,\quad P_\omega^+ \Omega=0~.
\end{align}
Hence on the localization locus we can choose a gauge where $\tilde \phi$ and $\varphi$ are both diagonal,
\begin{equation}
\tilde \phi= {\rm diag}({\tilde \phi}^a)~,\qquad \varphi={\rm diag }(\varphi^a)~,\qquad a=1,...,N-1~.
\end{equation}
Generically the gauge group $G$ is broken to its Cartan subgroup $H=U(1)^{N-1}$ and the path integral is over $H$ bundles \cite{Blau:1994rk,Blau:1995rs}, effectively reducing the problem to an abelian one. Topologically distinct $H$ bundles are distinguished by their fluxes $k_a^n\in {\mathbb Z}$ through a basis of two-cycles $\{ C^n \} $ in $M$,
\begin{equation}
\label{quant}
{1\over 2\pi}\int_{C^n} F^a= k^a_n~.
\end{equation}
Equation \eqref{locus} implies that the ${\tilde \phi}^a$ are constant while the $\varphi^a$ are invariant along $v$
\begin{equation}
d {\tilde \phi}^a=0~, \qquad \iota_v d\varphi^a=0~.
\end{equation}
Next we consider the localization conditions \eqref{locusa}. In principle there are many possible choices for  $\Omega$, here we will consider the following:
\begin{align}
\label{locdefomega}
\Omega=F-{2\over s+\tilde s}\,\star( \kappa \wedge d_A  \varphi)-\hat\Omega (\varphi)~,
\end{align}
where the middle term is related to (\ref{tran-ell-eqs}) (it arises naturally from the 5D perspective) and
 $\hat \Omega(\varphi)$ is not yet specified. With this $\Omega$ the $H$ field strength $F^a$  is completely determined away from the fixed points by \eqref{locusa}. For example with the canonical choice \eqref{canchoice} for the function $\omega$ defining the projector  $P^+_\omega$ we get
\begin{eqnarray}
 F^a &=&{s+\tilde s\over 2 ||v||^2} \star \big( \kappa \wedge d\varphi^a\big)-{1\over 2 ||v||^2} \kappa \wedge d\big((s-\tilde s)\varphi^a\big)+\\
&+&{\cos\omega \over 2 ||v||^2}  \star\Big(\kappa\wedge \big(2 \iota_v {\hat \Omega}^a+\varphi^a d (s-\tilde s)\big)\Big)+\Big({\hat \Omega}^a\! - {1\over ||v||^{2}} \kappa \wedge \iota_v {\hat \Omega}^a) \Big)~.
\end{eqnarray}
Finally we need to impose the Bianchi identity for  $F^a$. Because $s,\tilde s$ and $\varphi^a$ are all invariant along $v$ this results in a single scalar constraint:
\begin{equation}
\label{bianchired}
\star(\kappa \wedge d F^a)=0~.
\end{equation}
This is a second order differential equation for $\varphi^a$ whose detailed structure depends on the choice of $\hat \Omega$. As we will see below in some examples this equation may have a unique solution in every flux sector but we could not show that this is the case in general. A large class of cases is considered in detail in \cite{Fest-2018}.

The discussion above determines the localization locus away from the fixed points of $v$. At the fixed points where $\cos\omega=1$ the projector $P^+_\omega$ is the projector on self-dual two forms while at those where $\cos\omega=-1$ it reduces to the projector on anti-self-dual two forms. On top of the supersymmetric configurations described above we can then add point-like instantons at the fixed points with $\cos\omega=1$ and point-like anti-instantons at the fixed points with  $\cos\omega=-1$.

\subsubsection{$S^4$}
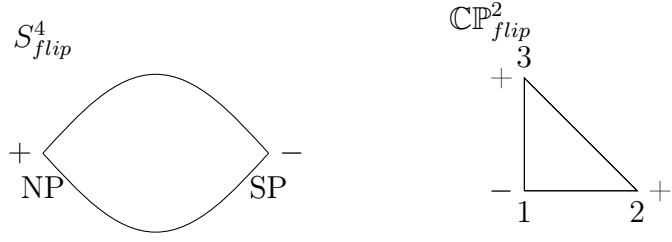
\begin{figure}
	\begin{center}
		\begin{tikzpicture}[scale=1.5]
		\node at (0,1){ $S^4_{flip}$ };
		\draw [-] (0,0) node[left] {{+}} sin (1,.7)  ;
		\draw [-] (1,.7) cos (2,0) node[right] {$-$};
		\draw [-] (0,0) node[left] {{+}} sin (1,-.7)  ;
		\draw [-] (1,-.7) cos (2,0);
		\node at (0,-.3) {NP};
		\node at (2,-.3) {SP};
		\end{tikzpicture}\qquad\qquad
		\begin{tikzpicture}[scale=1.5]
		\node at (-0.3, 1.5) {$\mathbb{CP}^2_{flip}$} ;
		\draw [-] (0,0) node[left] {{$-$}} -- (1,0) node[right] {{+}} -- (0,1) node[left] {{+}} -- (0,0);
		\draw [-] (0,0) node[below] {1} -- (1,0) node[below] {{2}} -- (0,1) node[above] {{3}} -- (0,0);
		\end{tikzpicture}\qquad
		\caption{The assignments of anti self-dual ($+$) and self-dual ($-$) instantons on $S^4$ and $\mathbb{CP}^2$ that we consider in our localization examples. }\label{fig_S4CP2}
	\end{center}
\end{figure}
As a first example we consider the manifold $S^4$ and determine the localization locus. To start we consider the round metric on $S^4$ . Explicit expressions for the metric and other objects in the coordinates we are using can be found in appendix \ref{app:S4}.
As in section \ref{exsf}, we take $v=\partial_\alpha+\partial_\beta$ while $s=2 \cos^2(\theta/2)$ and $\tilde s=2\sin^2(\theta/2)$ which correspond to Pestun's theory on the round $S^4$.

The localization term used in \cite{Pestun:2007rz} is
\begin{equation}
\Omega(\varphi)= \big(1+\cos^2\theta\big)^{1\over 2}\left(F-{2\over s+\tilde s}\,\star( \kappa \wedge d_A  \varphi)+{\cos\theta\over 2(1+\cos^2\theta)}\, \varphi \,d\kappa\right)~.
\end{equation}
This is of the form \eqref{locdefomega} apart from the  $(1+\cos^2\theta)^{1\over 2}$ prefactor whose presence does not modify the localization locus.
Proceeding as described above we determine $F^a$ away from the poles
\begin{equation}
\label{solp}
F^a ={1\over \sin^2\theta} \Big(\star \big( \kappa \wedge d\varphi^a\big)-\kappa \wedge d\big(\cos\theta\,\varphi^a\big)\Big)~.
\end{equation}
Imposing the Bianchi identity for $F^a$ then results in the following equation for $\varphi^a$
\begin{equation}
\label{lappe}
\nabla^2 \varphi^a=2\varphi^a~.
\end{equation}
The nonzero solutions to this equation are at least of order $\sin(\theta)^{-2}$ at one or both of the two poles. For this singular behavior the evaluation of the observable \eqref{obsdef} on the solution \eqref{solp} gives rise to integrals diverging at the poles. Moreover in order to prove that \eqref{obsdef} is supersymmetric one needs to perform integrations by parts that would not be allowed. For these reasons the only allowed solution to \eqref{lappe} is $\varphi^a=0$. Hence the localization locus away from the poles is
$$
\tilde \phi^a={\rm const}~,\quad \varphi^a=0~,\quad F^a=0~.
$$
We also have localized instantons at the north pole and anti-instantons at the south pole.

It is worth commenting on the relation with Pestun's treatment of the localization locus. On a round $S^4$ and with the choices made above we can make use of the special property \eqref{roundS4-dk} and formula (\ref{identity-FF-proj}) to  rewrite the bosonic terms in the localization Lagrangian as
\begin{align}
\label{ploc}
{\cal L}_{\rm{loc}}^B\mathrm{Vol}_{M} =& {\rm Tr}\!\left[2\cos^2\!\Big({\theta\over 2}\Big)\!\left(\!F^+\!-{\varphi \, d\kappa^+ \over 4 \cos^2(\theta/2)} \!\right)^{\!2}\!\!+ 2\sin^2\!\Big({\theta\over 2}\Big)\! \left(\!F^-\!+{\varphi \,d\kappa^-\over 4 \sin^2(\theta/2)} \!\right)^{\!2}\right.\cr&\quad+\left. (d_A\varphi)^2+(d_A\tilde \phi)^2- \mathrm{Vol}_{M} [\tilde\phi,\varphi]^2+4 F\wedge d_A (\varphi\kappa) \right ] ~ .
\end{align}
The last term above vanishes upon integrating by parts and we are left with a sum of squares. Hence away from the poles we obtain
\begin{equation}
d\tilde \phi^a=0~,\qquad d \varphi^a=0~,\qquad F^a={2\over \sin^2 \!y}\varphi^a \kappa\wedge d\cos y~.
\end{equation}
Enforcing Bianchi results in $\varphi^a=0$. This derivation implicitly assumes that integration by parts of the last term in \eqref{ploc} can be carried out. This excludes configurations for $\varphi^a$ that are singular at the poles.

We can deform the Killing vector away from $v=\partial_\alpha+\partial_\beta$. This can be accompanied by changes of the metric and other deformations as discussed in section \ref{secdefs}. For instance we may consider the theories on squashed $S^4_{\epsilon_1,\epsilon_2}$ described in \cite{ Hama:2012bg, Pestun:2014mja}. For generic $\epsilon_i$ it is no longer possible to rearrange the localization terms as in \eqref{ploc}, which makes the determination of the localization locus more subtle~\cite{Hama:2012bg}. Nevertheless, following the logic leading to \eqref{solp} we can argue that for small deformations around $v=\partial_\alpha+\partial_\beta$ the differential equation \eqref{lappe} for $\varphi^a$ is changed only slightly. Hence it continues to have only $\varphi^a=0$ as an acceptable solution.

\subsubsection{$\mathbb{CP}^2$}\label{CP2-subsection}

 As a second example we consider the manifold $\mathbb{CP}^2$ and we look in detail at the structure of the localization locus.
 We start by choosing the Fubini study metric. Expressions for the Fubini study metric and other quantities in the coordinates we use are collected in appendix \ref{app:CP2}.
We also make a choice of Killing vector field
\[
	v=2(\partial_\alpha+\partial_\beta) , \qquad \kappa =g(v)= 2(1-x-y)(x \, d\alpha + y\, d\beta).
 \]
The norm of $v$ vanishes along the line $x+y=1$ and not just at isolated points. This choice of $v$, however, makes the analysis below simpler and prepares the ground for us to consider a generic $v$. With these choices we can set
\[
s+\tilde s = 2, \qquad  \frac{s-\tilde s}{2}=\cos \omega =2x+2y-1 .
\]

There is one harmonic 2-form $\Omega^h$
\begin{equation}
\Omega^h=dx\wedge d\alpha+ dy\wedge d\beta~,
\end{equation}
which we will use to specify the localization term \eqref{locdefomega}
\begin{align}
\Omega=F- \star \big(\kappa \wedge D \varphi\big)+ \varphi \, \Omega^h~.
\end{align}
The localization equations \eqref{locus} result in the following expression for $F^a_{\mu\nu}$
\begin{eqnarray}
 F^a&=&{1\over 2 ||v||^2}\star\big( \kappa \wedge d\varphi^a\big)-{\cos \omega\over ||v||^2}\big( \kappa \wedge d\varphi^a\big )-\varphi^a \Omega^h~.
\end{eqnarray}
The Bianchi identity then gives rise to equation \eqref{bianchired} which reads
\begin{equation}
{1\over 2 }\left(\nabla^2 +4 x \partial_x +4 y \partial_y\right ) \varphi^a=0~.
\end{equation}
This equation is separable in the coordinates $x+y$ and ${x\over y}$. An analysis similar to that on $S^4$ shows that the only acceptable solution is a constant so that
\begin{equation}
F^a= k^a \Omega^h~,\qquad  \varphi^a=k^a\in Z ~ .
\end{equation}
The constants $k^a$ take values in the integers because the flux of $F$ is quantized as in \eqref{quant}.

Next we can modify our choice of $v$ to be
\begin{equation}
v=2(1+\epsilon) \partial_\alpha+2(1-\epsilon)\partial_\beta~,\qquad -{1\over 2}<\epsilon<{1\over 2}~.
\end{equation}
In order to keep the analysis of the localization locus simple it is convenient to change the metric by a nonsingular Weyl rescaling in order to be able to satisfy $s+\tilde s=2$. Because the Weyl rescaling of the metric is a $\delta$-exact deformation it will not change the value of supersymmetric observables. Hence we set the metric to be
\begin{equation}
ds^2= A(x,y)^2 ds^2_{FS}~,\qquad  A(x,y)={1\over \sqrt{1+ 4 \epsilon (1+\epsilon) x -4\epsilon (1-\epsilon) y }}~,
\end{equation}
where $ds^2_{FS}$ is the Fubini-Study metric \eqref{Fubmet}. We also take
\begin{equation}
\cos\omega=A(x,y) (2 x (1+\epsilon)+2 y (1-\epsilon)-1)~.
\end{equation}
For $-{1\over 2}<\epsilon<{1\over 2}$ the value of $\cos\omega$ at the fixed points is unchanged from the case considered above.
Finally we make the following choice for $\Omega$
\begin{align}
\Omega=F-\star \big(\kappa \wedge D \varphi\big)- \varphi\star \big(\kappa \wedge d \log(A) \big)+A \,\varphi\, \Omega^h~.
\end{align}
Proceeding as before \eqref{bianchired} results in an equation for $\varphi$ that is no longer separable. The equation is satisfied if $A \varphi $ is  a constant and this should be the only solution. The corresponding $F$ field strength is\begin{equation}
F^a= k^a \Omega^h~,\qquad \varphi^a=  A^{-1} k^a~, \qquad k^a \in Z~.
\end{equation}
Let us comment on certain properties of this solution that warrant further study.  Our solution can be written as
\bea
   F^a + \cos \omega ~\varphi^a = k^a \Omega^h + \left (2x (1+\epsilon) + 2y (1-\epsilon) -1 \right ) k^a \label{non-equiv-curv}
\eea
and it is annihilated by $(d+ \iota_v)$. It would be natural to say that it is the curvature of an equivariant line bundle.
However, the concrete solution (\ref{non-equiv-curv}) is not an element in the second integral equivariant cohomology on $\mathbb{CP}^2$.
 At this point we do not know how to interpret this conflict which requires further detailed study.

\subsection{The super-determinant}\label{sec_1loop}

Having found the localization locus, what remains is the Gaussian integral around it.
We have explained earlier that the Gaussian integral eventually produces (the square root of) a super-determinant.
Our entire computation is organized and goes along the same lines as the original  calculation of Pestun (see subsection 4.4 in \cite{Pestun:2007rz}).
However, it is instructive to formalize this computation in different terms. We will explain in a future paper \cite{Fest-2018} how ${\cal N}=2$ supersymmetry plus gauge fixing organizes all fields into a double complex, with $\gd$ acting vertically as an equivariant differential and a second differential $\gdh$ acting horizontally corresponding to the transversely elliptic complex in our  setup.
The superdeterminant is that of the operator $\gd^2=i {\cal L}_v + G_{a_0}$ taken over the cohomology of $\gdh$. Now we restrict to the trivial gauge background, i.e. all fields are zero except one scalar field $\tilde\phi$ having a vev $a_0$. As the gauge bundle is trivial, one can then write
\be
\sdet(i {\cal L}_v + G_{a_0})=\prod_{\ga\in{\textrm{roots}}}\sdet_{H_{\gdh}}\big(\bra a_0,\ga\ket+i{\cal L}_v\big).\nn
\ee
So one may well regard $a_0$ as a constant in the following computation.

To compute $\sdet_{H_{\gdh}}\big(a_0+i{\cal L}_v\big)$, one needs to compute $H_{\gdh}$ equivariantly, so that one can read off the weights of all the $U(1)$ actions and thereby write down the eigenvalues of  $i{\cal L}_v$. To summarize, our remaining task is to compute the equivariant cohomology ${H_{\gdh}}$ (in fact only the index is needed). What is the complex $(E^{\sbullet},\gdh)$ for which we want to compute $H_{\gdh}$?

\smallskip

\noindent {\bf Elliptic case}

Let us start from the simpler setup of  Donaldson-Witten theory where the corresponding complex is the instanton deformation complex already given in \eqref{elliptic-complex},
\be
0\to\Go^0(X)\to\Go^1(X)\to\Go^{2+}(X)\to0~,\label{inst_complex_4}
\ee
which is elliptic.
In the notation of the full cohomology complex \eqref{full-cohom-tran}, the three terms come from the fields $c,A,\chi$ respectively.
One also needs to add a two term complex $\Go^0\stackrel{\Gd}{\to}\Go^0$ represented by the fields $\varphi$ and $\eta$. But since $\Gd$ has zero index, one often ignores this last complex (one says that the two terms are 'cancelled'). In the non-equivariant case the index is easy to get with no computation
\be
\ind_{asd}=b^0-b^1+b^{2+}=\frac12(\chi+\tau)~,\nn
\ee
where $\chi$ is the Euler number and $\tau$ is the signature.
Including the gauge part, say for $SU(2)$, the index is
\be
\ind_{asd}=\frac32(\chi+\tau)-8k~,\nn
\ee
where $k$ is the instanton number.
Note that the dimension of the instanton (asd) moduli space is minus this index. For anti-instantons (sd), one flips the sign of $\tau$ and $k$.

For the equivariant Donaldson-Witten theory, one computes the cohomology of \eqref{inst_complex_4} equivariantly. We follow the method of \cite{AtiyahBott_Lef_I,AtiyahBott_Lef_II} that uses equivariant localization. We follow this route because it turns out that the index of the complex we need can be obtained by small tweaks of the same calculation around the torus fixed points.
A systematic treatment of such computations will be left to a future paper \cite{Fest-2018}.
According to \cite{AtiyahBott_Lef_I}
\be
\ind_{eq}=\sum_{p_i\in\textrm{fixed pt}}\frac{\chi_{eq}(E^{\sbullet}_{p_i})}{\det(1-df)}~,\label{AB}
\ee
where $\chi_{eq}(E^{\sbullet}_{p_i})$ stands for the equivariant Euler character of the fibre $E^{\sbullet}_{p_i}$ and $f:M\to M$ is the diffeomorphism induced by the group $G$ (in our case $U(1)^2$) action on $M$.

We start with $S^4$.
If there is an almost complex structure on $M$ then the complex \eqref{inst_complex_4} is isomorphic to the direct sum of  $(\Go^{0,\sbullet},\bar\partial)$ and its conjugate. But as \eqref{AB} only involves local data of $E^{\sbullet}$, we can still exploit such isomorphism to simplify computation even though $S^4$ is not almost complex.
We describe $S^4$ as the quaternion projective space $\BB{HP}^1$
\bea
S^4\simeq \{[q_1,q_2]|q_{1,2}\in\BB{H}\}/\sim~, ~~~\textrm{where}\,[q_1,q_2]\sim [q_1q,q_2q]~,~~q\in\BB{H}^*~.\nn
\eea
We use local inhomogeneous coordinate $q=q_1q_2^{-1}$ to cover the northern hemisphere and $q^{-1}$ for the southern hemisphere.
The $U(1)^2$ isometries act by \emph{left} multiplying $q_1\to sq_1$ and $q_2\to tq_2$, where we use the same letter $s,t$ for a phase as well as the character.
Writing $q=z+jw$ with $z,w\in\BB{C}$ gives us a local complex structure at the north pole. One reads off the action of $s,t$ on $z,w$ as $z\to st^{-1}z$, $w\to s^{-1}t^{-1}w$. Then at the north pole, one gets
\bea
 \chi_{eq}(\Go^{0,\sbullet})=(1-t/s)(1-st)~,~~~\det(1-df)=(1-s/t)(1-t/s)(1-s^{-1}t^{-1})(1-st)~.\nn
 \eea
So the local contribution to the index reads
\bea
 \textrm{north pole}:~~\frac{1}{(1-s/t)(1-s^{-1}t^{-1})}+\frac{1}{(1-t/s)(1-st)}~.\label{ind_north}
 \eea
At the south pole $q^{-1}=z'+jw'=(\bar z-jw)/|q|^2$ and so $\chi_{eq}(\Go^{0,\sbullet})=(1-s/t)(1-st)$, while $\det(1-df)$ does not change.
So the local contribution to the index is
\bea
 \textrm{south pole}:~~\frac{1}{(1-t/s)(1-s^{-1}t^{-1})}+\frac{1}{(1-s/t)(1-st)}~.\label{ind_south}
 \eea
Putting the two together one gets $\ind_{eq}=1$, which is obvious for the complex \eqref{inst_complex_4} for $S^4$.
Set $s=t=1$, the negative of the index times 3 (for $SU(2)$) gives us the expected dimension $-3$ of the instanton moduli space at a trivial background.

For the $\BB{CP}^2$ case, we use the inhomogeneous coordinates $[z_1,z_2,1]$ to cover one patch and we have torus action $z_1\to sz_1$ $z_2\to tz_2$. In other patches the actions are read off from the standard coordinate transformation. At the fixed points $z_1=z_2=0$ we have  $\chi_{eq}(\Go^{0,\sbullet})=(1-s^{-1})(1-t^{-1})$, while $\det(1-df)=(1-s)(1-s^{-1})(1-t)(1-t^{-1})$. The same expressions can be obtained in other patches
\bea
 \ind_{eq,asd}&=&\frac{1}{(1-s)(1-t)}+\frac{1}{(1-st^{-1})(1-t^{-1})}+\frac{1}{(1-s^{-1})(1-ts^{-1})}+c.c.=2~,\nn\\
\ind_{eq,sd}&=&\frac{1}{(1-s)(1-1/t)}+\frac{1}{(1-st^{-1})(1-t)}+\frac{1}{(1-s^{-1})(1-st^{-1})}+c.c.=1~.\nn
\eea
The difference between the two lines comes from conjugating one of the two local holomorphic coordinates at each fixed point, in order to turn $\Go^{2+}$ in \eqref{inst_complex_4} into $\Go^{2-}$.

The two indices of course agree trivially with the dimension of the instanton moduli space: $\dim_{\BB{CP}^2,\,asd}=8k-6$ and $\dim_{\BB{CP}^2,sd}=8k-3$,  with the factor of 3 coming from $SU(2)$ again.

\smallskip

\noindent {\bf Transversally elliptic case}

For our flipping instanton, the relevant complex to compute the equivariant index for is again the complex that controls the deformation \eqref{new-compex}
\bea
 \Go^0(X) ~\xrightarrow{d}~\Go^1 (X) \oplus \Go^0(X)~\xrightarrow{\tilde{D}}~P^+_\omega\Omega^{2} (X)\oplus \Go^0 (X)~,\nn\eea
with the expression for $\tilde D$ given by \eqref{expl-form-tildeD} (see Appendix \ref{App-TEP} for further discussion).
The flipping projector $P_{\go}^+$ approaches the self-dual projector when $\cos\go=1$ and anti self-dual when $\cos\go=-1$. These loci coincide with the torus fixed points.
In contrast to the previous elliptic case, one cannot cancel the two $\Go^0$ terms since they are crucial for transversal ellipticity. However, they can be cancelled at the torus fixed points, since at those loci $P^+_{\go}$ equals the asd or sd projector and $\tilde D$ is the direct sum $\tilde D=P^{\pm}d\oplus\Gd$ on the two summands. This means that as far as the localization computation of the index is concerned, one can obtain the new index from the computation done earlier plus some suitable modification to account for the asd/sd flips.

We start from the flipping instanton case of $S^4$ where the instantons/anti-instantons are at the north/south pole.
We introduce the notation $[1/(1-s)]^+$ to mean the expansion $1+s+s^2+\cdots$, and $[1/(1-s)]^-$ for $-s^{-1}-s^{-2}-\cdots$.
These two expansions are the equivariant characters of $H_{\bar\partial}^0(\BB{C})$ and $H_{\bar\partial}^1(\BB{C})$ respectively. The minus sign reflects the degree $H^1_{\bar\partial}$. The two different manners of rewriting $1/(1-s)$ into polynomials is related to deforming $\bar\partial$ with a vector field (coming from a group action) to trivialize the symbol \cite{Ellip_Ope_Cpct_Grp} (we only state this as a prescription, see the appendix of \cite{Pestun:2007rz} for a review).

We observe that the equations \eqref{ind_north}, \eqref{ind_south} can be thought of as the equivariant index of the cohomology of $(E^{\sbullet},\gdh)$ near the north/south pole, provided one expands the fraction into power series: we rewrite \eqref{ind_north}, \eqref{ind_south} using the new notation
\bea \textrm{north pole}:~~[1/(1-st^{-1})]^+[1/(1-s^{-1}t^{-1})]^++[1/(1-s^{-1}t)]^-[1/(1-st)]^-~,\label{ind_north_exp}\\
\textrm{south pole}:~~[1/(1-s^{-1}t)]^-[1/(1-s^{-1}t^{-1})]^++[1/(1-st^{-1})]^+[1/(1-st)]^-~.\label{ind_south_exp}\eea
The north/south pole 'know about each other' only through the choice of a vector field that trivializes the symbol of the complex.
Here the shift from $+$ to $-$ regularisation is attributable to the fact that $z',\,z$ have opposite weights, but $w',\,w$ have the same weights \footnote{Here and next we always make the technical assumption that one can deform the symbol with a vector field so that the symbol of the complex is trivialized except at torus fixed points. Otherwise one has to use the more general formula given in lecture 8 of \cite{Ellip_Ope_Cpct_Grp}.}.
Here one can check that the choice of $\pm$ at the two poles conspire so that, after expanding each term, everything cancels except one $s^0 t^0=1$, same as what we got earlier without the expansion. This is due to the fact that we were dealing with an elliptic complex before.

For the flipping instantons, the complex at the south pole is the sd complex (with projector $P^-$), following \cite{Pestun:2007rz} one flips the weight and regularisation at the south pole,
\bea
\textrm{north pole}:~~[1/(1-st^{-1})]^+[1/(1-s^{-1}t^{-1})]^++[1/(1-s^{-1}t)]^-[1/(1-st)]^-~,\label{ind_north_exp_F}\\
\textrm{south pole}:~~[1/(1-st^{-1})]^-[1/(1-s^{-1}t^{-1})]^-+[1/(1-s^{-1}t)]^+[1/(1-st)]^+~.\label{ind_south_exp_F}
\eea
Expanding everything using the given prescription and tracking down cancellations, the index is
\be
 \ind_{eq}=(\sum_{i,j\geq0}+\sum_{i,j\geq1})\big((st^{-1})^i(s^{-1}t^{-1})^j+(s^{-1}t)^i(st)^j\big)~.\label{ind_S4_flip}
\ee
It is no longer a finite polynomial, but a Laurent polynomial infinite in both directions, i.e. it is a formal expression and one cannot evaluate it at concrete values of $s,t$. To compare with \cite{Pestun:2007rz}, one \emph{formally} sets $st^{-1}=q=s^{-1}t^{-1}$
\be
\ind_{eq}\to2+\sum_{n\in\BB{Z}}2|n|q^n~.\nn
\ee
The contribution 2 are accounted for by the ghost zero modes.
Having obtained the equivariant index, the super determinant reads
\bea \sdet_{H_{\gdh}}\big(a_0+i{\cal L}_v\big)&=&
\prod_{i,j\geq0}\big(a_0+i\ep_1+j\ep_2\big)\prod_{i,j\geq1}\big(-a_0+i\ep_1+j\ep_2\big)\nn\\
&&\prod_{i,j\geq0}\big(-a_0+i\ep_1+j\ep_2\big)\prod_{i,j\geq1}\big(a_0+i\ep_1+j\ep_2\big)~,\nn\eea
where we have written $s=e^{\frac{i}{2}(\ep_2-\ep_1)}$, $t=e^{\frac{i}{2}(\ep_1+\ep_2)}$ and so for, say, for a term $s^2t^4$ in \eqref{ind_S4_flip}, one gets the $i{\cal L}_v$ eigen-value $-\ep_1-3\ep_2$.
Here the second line corresponds to \eqref{ind_S4_flip}.

\begin{figure}
\begin{center}
\begin{tikzpicture}[scale=1.5]
\node at (0.5,1.4) {$\mathbb{CP}^2_{asd}$ } ;
\draw [-] (0,0) node[left] {{+}} -- (1,0) node[right] {{+}} -- (0,1) node[left] {{+}} -- (0,0);
\draw [-] (0,0) node[below] {1} -- (1,0) node[below] {{2}} -- (0,1) node[above] {{3}} -- (0,0);
\end{tikzpicture}\qquad
\begin{tikzpicture}[scale=1.5]
\node at (0.5,1.4) {$\mathbb{CP}^2_{flip}$ } ;
\draw [-] (0,0) node[left] {{-}} -- (1,0) node[right] {{+}} -- (0,1) node[left] {{+}} -- (0,0);
\draw [-] (0,0) node[below] {1} -- (1,0) node[below] {{2}} -- (0,1) node[above] {{3}} -- (0,0);
\end{tikzpicture}\qquad
\begin{tikzpicture}[scale=1.5]
\node at (0.5,1.4) {$\mathbb{CP}^2_{flip'}$ } ;
\draw [-] (0,0) node[left] {{+}} -- (1,0) node[right] {{-}} -- (0,1) node[left] {{-}} -- (0,0);
\draw [-] (0,0) node[below] {1} -- (1,0) node[below] {{2}} -- (0,1) node[above] {{3}} -- (0,0);
\end{tikzpicture}
\caption{Three distributions of asd/sd instantons on $\BB{CP}^2$, where $+$ corresponds to anti self-dual, $-$ to self-dual. At corner 1, we have inhomogeneous coordinates $[z_1,z_2,1]$, corner 2 $[1,z_2/z_1,1/z_1]$ and corner 3 $[z_1/z_2,1,1/z_2]$.}\label{fig_CP2}
\end{center}
\end{figure}
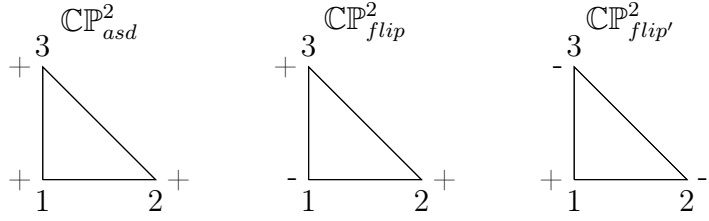

Now we will modify the regularisation scheme at each fixed point to get the index for the new instantons.
For $\BB{CP}^2$ one can write
\be \ind_{\BB{CP}^2,asd}=\big[\frac{1}{1-s}\big]^+\big[\frac{1}{1-t}\big]^++\big[\frac{1}{1-1/s}\big]^-\big[\frac{1}{1-t/s}\big]^-
+\big[\frac{1}{1-s/t}\big]^+\big[\frac{1}{1-1/t}\big]^-+c.c.\label{CP2_loc}
\ee
where the three terms come from the three corners of figure \ref{fig_CP2} (and the local coordinates are also labeled there).
All terms cancel except $2s^0t^0=2$ as before. We now consider the new instanton complex that approaches asd at the second and third fixed points and sd at the first one as in the middle panel of figure \ref{fig_CP2}. The modification one does to \eqref{CP2_loc} is to reverse at corner 2,3 the weights of $z_3$
\be
 \ind_{\BB{CP}^2,flip}=\big[\frac{1}{1-s}\big]^+\big[\frac{1}{1-t}\big]^++\big[\frac{1}{1-{\color{blue}s}}\big]^-\big[\frac{1}{1-t/s}\big]^-
+\big[\frac{1}{1-s/t}\big]^+\big[\frac{1}{1-{\color{blue}t}}\big]^-+c.c.\nn
\ee
This prescription is derived from regarding the flipping instanton as the reduction of 5D contact instantons on $S^5$ along the $U(1)$ of weight $[1,1,-1]$, and use the 5D index calculations.
It is yet unclear how this prescription compares to the one used in \cite{Pestun:2007rz}, in particular, it is desirable to have a purely 4D treatment of the index calculation.

Tracking down the cancellations one gets the $\ind_{\BB{CP}^2,flip}$ as the following graph plus its complex conjugate ($s,t\to s^{-1},t^{-1}$).
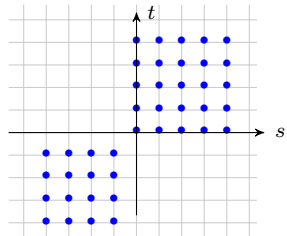
\begin{center}
\begin{tikzpicture}[scale=1]
\draw [step=0.3,thin,gray!40] (-.2,-1.4) grid (3.1,1.7);

\foreach \x in {0,1,...,4}
\foreach \y in {0,1,...,4}
\node at (1.5+0.3*\x,0.3*\y) [blue] {$\sbullet$};

\foreach \x in {0,1,...,3}
\foreach \y in {0,1,...,3}
\node at (1.2-0.3*\x,-0.3-0.3*\y) [blue] {$\sbullet$};

\draw [->] (1.5,-1.1) -- (1.5,1.6) node [right] {\scriptsize{$t$}};
\draw [->] (-.2,0) -- (3.2,0) node [right] {\scriptsize{$s$}};

\end{tikzpicture}
\end{center}
That this picture formally looks the same as $S^4$ means nothing, since the $s,t$ parameters will be identified differently.

To compute the index for the third configuration in figure \ref{fig_CP2}, we regard the flipping instanton as the reduction from $S^5$ along a $U(1)$ of weight $[-1,-1,1]$ (the miss/alignment of $[-1,-1,1]$ with the standard Hopf vector field $[1,1,1]$ tells us whether it is asd or sd at each corner, see \cite{Fest-2018}).
Correspondingly
\be \ind_{\BB{C}P^2,flip'}=\big[\frac{1}{1-{\color{blue}1/s}}\big]^+\big[\frac{1}{1-{\color{blue}1/t}}\big]^++\big[\frac{1}{1-1/s}\big]^-\big[\frac{1}{1-{\color{blue}s/t}}\big]^-
+\big[\frac{1}{1-{\color{blue}t/s}}\big]^+\big[\frac{1}{1-1/t}\big]^-+c.c.\nn
\ee
After cancellations, we get the same picture as the above with $s,t$ reversed, but as we also add the conjugate graph, the result is not changed.

As another example, we also give the result for the Hirzebruch surface $\BB{F}_1$,
\begin{figure}
\begin{center}
\begin{tikzpicture}[scale=1.3]
\draw [-] (0,0) node[left] {{+}} -- (1,0) node[right] {{-}} -- (1,2) node[right] {{-}} -- (0,1) node[left] {{+}} -- (0,0);
\node at (.5,-0.2) {\scriptsize{$2$}};
\node at (1.2,1) {\scriptsize{$3$}};
\node at (.5,1.7) {\scriptsize{$4$}};
\node at (-0.2,0.5) {\scriptsize{$1$}};
\end{tikzpicture}\qquad
\begin{tikzpicture}[scale=1]
\draw [step=0.3,thin,gray!40] (-.2,-1.4) grid (3.1,1.7);

\node at (-1.2,0) {\scriptsize$\ind_{\BB{F}_1,flip}=$};

\foreach \x in {0,1,...,4}
\foreach \y in {0,...,\x}
\node at (1.5+0.3*\x,0.3*\y) [blue] {$\sbullet$};

\foreach \x in {0,...,3}
\foreach \y in {0,...,\x}
\node at (0.9-0.3*\x,-0.3-0.3*\y) [blue] {$\sbullet$};

\draw [->] (1.5,-1.1) -- (1.5,1.6) node [right] {\scriptsize{$t$}};
\draw [->] (-.2,0) -- (3.2,0) node [right] {\scriptsize{$s$}};

\end{tikzpicture}
\end{center}
\caption{One distribution of $\pm$'s for the Hirzebruch surface $\BB{F}_1$ and the corresponding result of the index computation. }\label{fig_F2}
\end{figure}
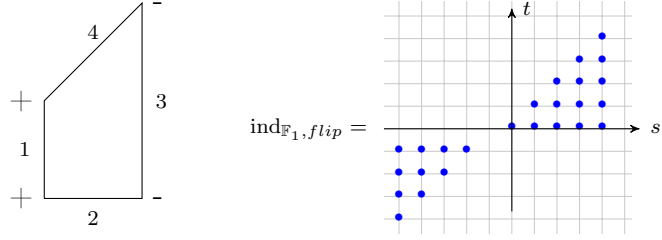
where $\BB{F}_1$ is realised as $S^2$ fibered over $S^2$ with degree 1, i.e.
\be
 \BB{F}_1\simeq\{[z_1,z_2;u_1,u_2]\}/\sim,~~~[z_1,z_2;u_1,u_2]\sim [\gl z_1,\gl z_2;\gl \mu u_1,\mu u_2],~~\gl,\mu\in\BB{C}^*.\nn
\ee
The $U(1)$'s $s$ and $t$ rotates the phase of $z_1,u_1$ respectively.
The assignment of asd/sd instantons and the corresponding result of the index computation is shown in figure \ref{fig_F2}.
In the three examples above, setting $s=e^{-i\ep_1}$ and $t=e^{-i\ep_2}$, one can write down
the superdeterminant in a uniform manner.
\bea \sdet_{H_{\gdh}}\big(a_0+i{\cal L}_v\big)&=&\prod_{(p,q)\in C\cap\BB{Z}^2}\big(a_0+\ep_1p+\ep_2q\big)\prod_{(p,q)\in C^{\circ}\cap\BB{Z}^2}\big(-a_0+\ep_1p+\ep_2q\big)\nn\\
&&\prod_{(p,q)\in C\cap\BB{Z}^2}\big(-a_0+\ep_1p+\ep_2q\big)\prod_{(p,q)\in C^{\circ}\cap\BB{Z}^2}\big(a_0+\ep_1p+\ep_2q\big)~.\label{infinite_prod}\eea
Here $C$ is a rational cone and $C^{\circ}$ is its interior.

Since the structure is rather universal, it is useful to define
\be \label{eq:Upsilonfunc}
 \Upsilon^C(x|\ep_1,\ep_2)=\prod_{(p,q)\in C\cap\BB{Z}^2}\big(x+\ep_1p+\ep_2q\big)\prod_{(p,q)\in C^{\circ}\cap\BB{Z}^2}\big(-x+\ep_1p+\ep_2q\big)~.\nn
 \ee
Then the superdeterminant in all cases, complete with the Lie algebra factor reads
\be
\sdet(i {\cal L}_v + G_{a_0})=\prod_{\ga\in{\textrm{roots}}}\Upsilon^C(\bra a_0,\ga\ket|\ep_1,\ep_2)\Upsilon^C(-\bra a_0,\ga\ket|\ep_1,\ep_2)~.\nn
\ee

When one computes the index by summing up local contributions from toric fixed points, one generally gets a Laurent polynomial infinite in both directions. In the above examples, the results organize into a cone and its negative, and so by reversing the negative cone, one can in fact evaluate the infinite product \eqref{infinite_prod} provided $\ep_1,\ep_2$ is within the dual cone. That is for $\vec n\in C$, $\vec n\cdotp \vec \ep>0$ and increases to infinity as $|\vec n|\to\infty$. Thus one can use the zeta function to regulate the infinite product.
The $\Upsilon$-function defined in \eqref{eq:Upsilonfunc} is closely related to the generalized triple sine \cite{Winding:2016wpw} that appears in the 5D 1-loop computation on toric manifolds.

Given an arbitrary assignment of asd/sd at the toric fixed points, we currently do not know how to combine the local contributions and so we cannot write down the 1-loop part of the partition function.

Secondly the parameters $\ep_{1,2}$ are real, but one can factorize the $\Upsilon$ function into as many factors as there are toric fixed points, provided one gives a non-zero imaginary part to $\ep_{1,2}$.
This factorization led to the conjecture that the full partition function is glued from copies of partition functions of $\BB{C}^2_{eq}$, there being one copy for each toric fixed point. Except for taking into account the non-zero $c_1$ of the gauge bundle, the Coulomb branch parameter $a_0$ is shifted between the different copies of the local contribution.

\subsection{The full answer}\label{ss:full-answer}

Let us summarize here what we think about the general answer. Before doing this let us comment on the analytical properties of instanton and anti-instanton partition functions on $\mathbb{C}^2$. With the complex coordinates $(z_1, z_2) \in \mathbb{C}^2$ the change $z_1 \rightarrow z_1,  z_2 \rightarrow \bar{z}_2$ (or  $z_2 \rightarrow z_2,  z_1 \rightarrow \bar{z}_1$) will induce the map from instanton equations to anti-instanton equations. Therefore we have the following relations
\bea
Z_{\epsilon_1, \epsilon_2}^{\rm anti-inst} (a, \bar{q}) = Z_{\epsilon_1, -\epsilon_2}^{\rm inst} (a, \bar{q}) =
Z_{- \epsilon_1, \epsilon_2}^{\rm inst} (a, \bar{q})~,
\eea
where $\bar{q}$ is an instanton counting parameter. We stress that for complex $(\epsilon_1, \epsilon_2)$ the instanton and anti-instanton partition functions are not related by simple complex conjugation
\bea
Z_{\epsilon_1, \epsilon_2}^{\rm anti-inst} (a, \bar{q}) \neq \overline{ Z_{\epsilon_1, \epsilon_2}^{\rm inst} (a, q)}~.
\eea
The answer for $S^4_{\epsilon_1, \epsilon_2}$ with complex  $(\epsilon_1, \epsilon_2)$ is given by
\bea
Z_{S_{\epsilon_1, \epsilon_2}^4} = \int da~e^{S_{cl}} ~Z^{\rm inst}_{\epsilon_1, \epsilon_2} (ia, q)   Z^{\rm anti-inst}_{\epsilon_1, - \epsilon_2} (ia, \bar{q}) =
 \int da~e^{S_{cl}} ~Z^{\rm inst}_{\epsilon_1, \epsilon_2} (ia, q)   Z^{\rm inst}_{\epsilon_1, \epsilon_2} (ia, \bar{q})
\eea
and it is holomorphic in $(\epsilon_1, \epsilon_2)$.
Only for real $\epsilon$'s the original formula by Pestun holds
\bea
Z_{S_{\epsilon_1, \epsilon_2}^4} = \int da~e^{S_{cl}} ~|Z^{\rm inst}_{\epsilon_1, \epsilon_2} (ia, q)|^2 ~.
\eea
The general answer for $Z_{M_{\epsilon_1, \epsilon_2}}$ with $p$ plus points and $(l-p)$ minus points will be given by
\bea
\sum\limits_{{\rm\small{discrete~} k_i}}
~~\int\limits_{\bf h} da~ e^{-S_{cl}} \prod\limits_{i=1}^{p} Z^{\rm inst}_{\epsilon_1^i, \epsilon_2^i} \Big (ia + k_i (\epsilon_1^i, \epsilon_2^i), q \Big )  \prod\limits_{i=p+1}^l Z^{\rm anti-inst}_{\epsilon_1^i, \epsilon_2^i} \Big (ia + k_i  (\epsilon_1^i, \epsilon_2^i), \bar{q} \Big ) ~. \label{final-answer-sum}
\eea
 The parameters  $(\epsilon_1^i, \epsilon_2^i)$ can be defined from $T^2$-action around the fixed point $x_i$. The classical action can be evaluated by the localization of the observable (\ref{obsdef}) and for this we need to know the values of $\phi$ at the fixed points (in particular $k_i(\epsilon_1^i, \epsilon_2^i)$).
The main problem is to fix the functions $k_i(\epsilon_1^i, \epsilon_2^i)$ which corresponds to fluxes. The original idea from \cite{MR2227881} is to declare $(F^a + \cos \omega ~\varphi^a)$  (we look at this combination along Cartan sub-algebra ${\mathbf h}$) to be the curvature of an equivariant line bundle and thus to be an element of the integral equivariant cohomology class. If we accept this approach then the shift functions can be written as follows (more details in \cite{Fest-2018})
\bea
k_i  (\epsilon_1^i, \epsilon_2^i) = k_i \epsilon^i_1 + k_{i-1} \epsilon^i_2~,\label{shift-equivariant}
\eea
where $k_i$ is vector composed from integers (the number of components of this vector is given by dimension of Cartan subalgebra).  The sum in
the final answer (\ref{final-answer-sum}) is taken over subset of the integers. This logic is common and a natural belief (for recent discussions see e.g. \cite{Bershtein:2015xfa, Bershtein:2016mxz,Hosseini:2018uzp, Crichigno:2018adf}). In section \ref{CP2-subsection} we have tried to analyze the concrete localization locus for the concrete example of $\mathbb{CP}^2$ assuming only the integrality of $F$. This analysis results in shifts that are not compatible with the expression (\ref{shift-equivariant}).
 At the moment we are unsure how to interpret this observation and we leave the problem to further study.

\section{Summary}\label{s:summary}

In this work we have constructed an ${\cal N}=2$ supersymmetric gauge theory on a manifold which admits a Killing vector field with isolated fixed points.
We gave a cohomological description of this theory and thus explained the relation between Donaldson-Witten theory and Pestun's calculation on $S^4$.
In the follow up work \cite{Fest-2018} we study further the formal aspects of this cohomological theory and we stress the relation between supersymmetry and a transversely elliptic complex.

In this work we have conjectured the answer for the partition function for the general case. The main remaining challenge is  to prove  this answer and better understand how it arises.
It will be crucial to understand which geometrical data controls the fluxes, and to further refine the arguments presented in section \ref{s:localization}.
In  \cite{Fest-2018}  we are able to  be more precise about the general answer when the theory arises from the reduction of a supersymmetric 5D gauge theory.  The 5D language appears to be very powerful for repacking the 4D answer and it allows to fix some ambiguities in 4D. Its drawback is that it does not apply to all possible 4D theories.

It would be interesting to study what are the physical ramifications of our present construction. For instance our construction could be interpreted from the point of view of the AGT-paradigm \cite{Alday:2009aq}. It is also interesting to study how our formalism relates to the ``cutting and gluing'' formalism for supersymmetric path integrals described in \cite{Dedushenko:2018aox, Dedushenko:2018tgx}.

\bigskip
{\bf Acknowledgements:} We thank Nikita Nekrasov and Vasily Pestun for discussions.
The work of GF is supported by the ERC STG grant 639220.
The work of MZ  is supported in part by Vetenskapsr\r{a}det under grant \#2014-5517, by the STINT grant, and by the grant  ``Geometry and Physics"  from the Knut and Alice Wallenberg foundation.

 \appendix

\section{Conventions} \label{app:conventions}

Here we collect various relevant formulas and a summary of our conventions. They are based on those of \cite{WessBagger}, adapted to Euclidean signature.

\subsection{Flat Euclidean Space}

The metric is~$\delta_{\mu\nu}$, where~$\mu, \nu = 1, \ldots, 4$. The totally antisymmetric Levi-Civita symbol is~$\ep_{1234} = 1$. The rotation group is~$SO(4) = SU(2)_+ \times SU(2)_-$. A left-handed spinor~$\zeta_\alpha$ is an~$SU(2)_+$ doublet and has un-dotted indices. A Right-handed spinor $\bar \zeta_{\dot\alpha}$ is a doublet under~$SU(2)_-$. It carries a bar and dotted indices. In Euclidean signature, $\zeta$ and~$\bar \zeta$ are independent spinors as  $SU(2)_+$ and~$SU(2)_-$ are not related by complex conjugation. Dotted and undotted indices are raised acting on the left with the totally antisymmetric $2 \times 2$ matrix $\epsilon$ defined by $\epsilon^{12}=1$. Hence we have $\zeta^\alpha=\epsilon^{\alpha \beta}\zeta_\beta$ and $\bar \zeta^\alphadot=\epsilon^{\alphadot \betadot}\bar \zeta_\betadot $.
We write the~$SU(2)_+$ invariant inner product of~$\zeta$ and~$\eta$ as~$\zeta \eta=\zeta^\alpha \eta_\alpha$. Similarly, the~$SU(2)_-$ invariant inner product of~$\bar \zeta$ and~$\bar \eta$ is given by~${\bar \zeta} \bar \eta={\bar \zeta}_{\alphadot} \bar \eta^\alphadot$.

The sigma matrices are written in terms of the Pauli matrices~$\vec{\sigma} = (\sigma^1, \sigma^2, \sigma^3)$
\be
\label{sigmamat}
\sigma^\mu_{\alpha{\dot\alpha}} = (\vec{\sigma}, -i)~,\qquad \bar \sigma^{\mu{\dot\alpha} \alpha} = (-\vec{\sigma}, -i)~.
\ee
In Euclidean signature~$\sigma^\mu$ and~$\bar \sigma^\mu$ are not related by complex conjugation. The sigma matrices \eqref{sigmamat} satisfy the identities
\be
\sigma_\mu\bar \sigma_\nu + \sigma_\nu \bar \sigma_\mu = -2\delta_{\mu\nu}~, \qquad \bar \sigma_\mu \sigma_\nu + \bar \sigma_\nu \sigma_\mu = -2\delta_{\mu\nu}~.
\ee
We define the antisymmetric matrices
\be
\sigma_{\mu\nu} = {1 \over 4} (\sigma_\mu \bar \sigma_\nu - \sigma_\nu\bar \sigma_\mu)~, \qquad \bar \sigma_{\mu\nu} = {1 \over 4} (\bar \sigma_\mu \sigma_\nu - \bar \sigma_\nu \sigma_\mu)~.
\ee
They are self-dual and anti-self-dual respectively,
\be
{1\over 2} \epsilon_{\mu\nu\rho\lambda} \sigma^{\rho \lambda } = \sigma_{\mu\nu}~, \qquad {1\over 2} \epsilon_{\mu\nu\rho\lambda} \bar \sigma^{\rho\lambda} = - \bar \sigma_{\mu\nu}~.
\ee
Given a two form $\omega$ we can separate its $(2,0)$ and $(0,2)$ components as follows:
\be
\omega^+_{\alpha \beta}={1\over 2} \omega_{\mu\nu}\sigma^{\mu\nu}_{\alpha \beta}~,\qquad \omega^-_{\alphadot \betadot}={1\over 2} \omega_{\mu\nu}\bar \sigma^{\mu\nu}_{\alphadot \betadot}~.
\ee

\subsection{Differential Geometry}

We use Greek letters~$\mu, \nu, \ldots$ to denote curved indices and Latin letters~$a, b, \ldots$ to denote frame indices. Given a Riemannian metric~$g_{\mu\nu}$, we can define an orthonormal tetrad~${e^a}_\mu$. We denote the Levi-Civita connection by~$\nabla_\mu$. The corresponding spin connection is given by
\be
{\omega_{\mu a}}^b = {e^b}_\nu \nabla_\mu {e_a}^\nu~.
\ee
The covariant derivatives of the spinors~$\zeta$ and~$\bar \zeta$ are given by
\be
\nabla_\mu \zeta = \partial_\mu \zeta + {1\over 2}\omega_{\mu a b } \sigma^{ab} \zeta~, \qquad \nabla_\mu \bar \zeta = \partial_\mu \bar \zeta +  {1\over 2} \omega_{\mu a b } {\bar \sigma}^{ab} \bar \zeta~.
\ee
The Riemann tensor tis
\be
{R_{\mu\nu a}}^b = \partial_\mu {\omega_{\nu a}}^b - \partial_\nu {\omega_{\mu a}}^b + {\omega_{\nu a}}^c {\omega_{\mu c}}^b - {\omega_{\mu a}}^c {\omega_{\nu c}}^b~.\ee
The Ricci tensor is~$R_{\mu\nu} = {R_{\mu\rho\nu}}^\rho$, and~$R = {R_\mu}^\mu$ is the Ricci scalar. Note that, with these conventions, a round sphere has negative Ricci scalar.

\subsection{Differential forms}
We use the following conventions when switching to index free notation for differential forms.  An $n$-form is given by $\omega^{(n)}$:
\begin{equation}
\omega^{(n)}={1\over n!} \omega_{\mu_1\mu_2...\mu_n} dx^{\mu_1}\wedge dx^{\mu_2}\wedge...\wedge dx^{\mu_n}~.
\end{equation}
The Hodge star operator acts as follows:
\begin{equation}
(\star \omega)^{(4-n)}_{\mu_1...\mu_{4-n}}={1\over n!} \,\omega^{(n)}_{\nu_1...\nu_n}{\epsilon^{\nu_1...\nu_n}}_{\mu_1...\mu_{4-n}}~.
\end{equation}
It satisfies $\star\!\star\!\omega^{(n)}=(-1)^n \omega^{(n)}~.$

Given a vector field $v=v^\mu {\partial\over \partial x^\mu}$ the contraction of a differential $n$-form $\omega^{(n)}$ with $v$ is
\begin{equation}
(\iota_v \omega)^{(n-1)}_{\mu_1...\mu_{n-1}}=v^\nu \omega^{(n)}_{\nu\mu_1...\mu_{n-1}}~.
\end{equation}
Consider the $1$-form $\kappa$ obtained from $v$ via the metric
\begin{equation}
\kappa_\mu=g_{\mu\nu} v^\nu~,
\end{equation}
then the following relation holds
\begin{equation}
\iota_v\star \omega^{(n)}= (-1)^n \star(\kappa\wedge\omega^{(n)})~.
\end{equation}

\subsection{Other conventions}
We denote symmetrization over indices using round brackets,
\be
A_{(i_1\ldots i_p)}={1\over p!} \sum_{\sigma \in S_p} A_{i_{\sigma(1)}\ldots i_{\sigma(p)}}~.
\ee
Anti-symmetrization is denoted by square brackets
\be
A_{[i_1\ldots i_p]}={1\over p!} \sum_{\sigma \in S_p} (-1)^\sigma A_{i_{\sigma(1)}\ldots i_{\sigma(p)}}~.
\ee

\section{Examples arising from specific four-manifolds}\label{app:explicitgeom}

\subsection{$S^4$}\label{app:S4}

Here we collect the basic conventions we follow when using $S^4$ as an example. Let us introduce  the
coordinates $\theta, x,\alpha,\beta$ on $S^4$ where $\theta\in [0,\pi], \ x \in [0,1]$ and $\alpha,\beta\in [0,2\pi]$ are angles.
The round metric in these coordinates takes the form
\be
ds^2 = d\theta^2 + \sin^2\theta \left ( \frac{dx}{4x(1-x) } + x d\alpha^2 + (1-x) d\beta^2 \right )~ .
\ee
These coordinates allow us to think of $S^4$ as a degenerating $T^2$ fibration (the two angles spanning $T^2$) over the 2d surface with boundaries spanned by $\theta,x$, see figure \ref{fig_S4CP2}.
The cycle parametrized by $\alpha$ shrinks when $x=0$, and the $\beta$-cycle shrinks when $x=1$.
The north pole corresponds to $\theta = 0$ and the south pole to $\theta = \pi$.

To compare with the work of Pestun we choose the Killing vector field $v$ and its dual 1-form $\kappa$ as
\be
\label{chpesa}
v = \partial_\alpha + \partial_\beta , \qquad \kappa=g(v) = \sin^2\theta (x d\alpha  +  (1-x) d\beta ).
\ee
We also select $\cos(\omega)=\cos \theta$, which goes from $+1$ on the north pole to $-1$ on the south pole.
The projector on 2-forms we introduced in section 2 is
\be
\begin{split}
	P^+_{\theta} & = \frac{1}{1+\cos^2\theta} (1+\cos\theta \star - \sin^2\theta \; \hat\kappa\wedge \iota_v)  \\
	&=\frac{1}{1+\cos^2\theta} (1+\cos\theta \star - \kappa\wedge \iota_v)~ ,
\end{split}
\ee
which is everywhere smooth since $\kappa$ and $v$ are globally well defined.
Since $v$ and $\kappa$ both go to zero at the poles, we can see that $P^+_{\theta}$ collapses to the self-dual projector over the north pole and and to the anti-self-dual projector over the south pole. For the specific choice \eqref{chpesa} one can check that $d\kappa$ satisfies the following property
\be
P^+_{\theta} d \kappa = d\kappa~.\label{roundS4-dk}
\ee
The volume form in these coordinates is
\be
\mathrm{Vol}_{S^4} = \frac 1 2 \sin^3 \theta \; d\theta\wedge dx \wedge d\alpha \wedge d\beta .
\ee

\subsection{$\mathbb{CP}^2$} \label{app:CP2}

We introduce the following coordinate system on $\mathbb{CP}^2$. Two of the coordinates $x,y$ parametrize a simplex in $\mathbb{R}^2$,
\begin{equation}
x\geq0~,\qquad y\geq0~,\qquad x+y\leq 1~,
\end{equation}
see the right part of figure \ref{fig_S4CP2}.
On any point of the simplex $(x,y)$ there is a torus parametrized by two angles $\alpha$ and $\beta$. The 1-cycle on the torus parametrized by $\alpha$ shrinks to zero for $x=0$, the one parametrized by $\beta$ shrinks to zero for $y=0$ while the 1-cycle at constant $\alpha-\beta$ shrinks on the remaining boundary of the simplex $x+y=1$.
The relation of these coordinates to the inhomogeneous coordinates $[z_1,z_2,1]$ used in section \ref{sec_1loop} is as follows:
\begin{equation}
z_1=\left( {x\over 1-x-y}\right)^{1\over 2}e^{i\alpha}~,\quad  z_2=\left( {y\over 1-x-y}\right)^{1\over 2}e^{i\beta}~.
\end{equation}

There are three fixed points of the torus action at the corners of the simplex
\begin{equation}
1: (x=0,y=0)~,\quad 2: (x=1,y=0)~,\quad 3: (x=0,y=1)~.
\end{equation}
Consider the Fubini-study metric on $\mathbb{CP}^2$, which in the coordinates we just introduced is given by
\begin{equation}
\label{Fubmet}
ds_{FS}^2= {dx^2\over 4 x}+ { (dx +dy)^2\over 4(1\!-\!x\!-\!y)}+ {dy^2\over 4 y}+(1-x) x d\alpha^2 -2 x y d\alpha d\beta +(1-y) y d\beta^2~.
\end{equation}
The volume form in these coordinates is
\be
\mathrm{Vol}_{\mathbb{CP}^2} = \frac 1 4 dx \wedge dy \wedge d\alpha\wedge d\beta .
\ee
Next consider the Killing vector field
\[
v=2(\partial_\alpha+\partial_\beta) , \qquad \kappa =g(v)= 2(1-x-y)(x \, d\alpha + y\, d\beta).
\]
The norm of $v$ is
\[
||v||^2 = 4 (1 - x - y) (x + y).
\]
This  $v$ does not vanish at isolated points because $||v||^2=0$ for $x+y=1$; however, it is a useful starting point to analyze more general choices. With these choices we can set
\[
s+\tilde s = 2, \qquad
h=\frac{s-\tilde s}{2}=\cos\omega=2x+2y-1 ,
\]
so that $\cos^2 \omega=1-||v||^2$.

The projector $P^+_{\go}$ for these choices takes the form
\be
P_\go^+ = \frac{1}{1+(2x+2y-1)^2} \left ( 1 + (2x+2y-1)\star + 4(1-x-y)(x\, d\alpha+y\, d\beta) \wedge \iota_{\partial_\alpha+\partial_\beta} \right ) ,
\ee
which is nowhere singular and goes to $P^-$ at fixed point $1$ and to $P^+$  along the line $x+y=1$.
Note that $\cos\omega$ is equal to $-1$ at fixed point $1$ and is equal to $+1$ at fixed points $2$ and $3$. Hence the cohomological theory built using $P_\go^+$ in section \ref{subs:cohom-complex} would not be equivalent to the equivariant topological twist on $\mathbb{CP}^2$.

\section{Transverse elliptic problems in 4D and 5D}\label{App-TEP}

In this appendix we collect explicit expressions for the transversally elliptic problems which appear in the context of gauge theories in 4D and 5D. Since we are interested in symbols of operators, we consider only the case of an abelian theory and expand around the trivial connection.  All our geometrical considerations are local in nature since ellipticity and transversal  ellipticity are local notions.

Let us start from anti self-duality in 4D, $P^+ F = 0$. If we add the gauge fixing condition we obtain the following system of PDEs
\bea
&& d^\dagger A=0~,\nn \\
&& F_{12} + F_{34}=0~,\nn\\
&& F_{13} + F_{42}=0~,\\
&& F_{23} + F_{14}=0~.\nn
\eea
The symbol of the corresponding operator is
\bea
\sigma_{4D}^{sd} = \left (
\begin{array}{cccc}
	p_1 & p_2 & p_3 & p_4 \\
	-p_2 & p_1 & - p_4 &  p_3 \\
	-p_3 &  p_4 & p_1 & - p_2 \\
	-p_4 & - p_3 &  p_2 & p_1
\end{array}\right  )~,
\eea
and this is an elliptic symbol. If we add a fifth direction as a trivial circle and label this direction by zero then we can lift this elliptic problem to a transversely elliptic one (with respect to new $U(1)$ circle) simply by declaring that $A_0=0$.  However, we would like to rewrite this system in 5D gauge covariant terms
\bea
&& \partial^1 F_{01} +  \partial^2 F_{02} +  \partial^3 F_{03}+  \partial^4 F_{04}=0 ~, \nn\\
&& d^\dagger A=0~,\nn \\
&& F_{12} + F_{34}=0~,\label{5D-expl-system}\\
&& F_{13} + F_{42}=0~,\nn \\
&& F_{23} + F_{14}=0~,\nn
\eea
where now the Lorentz gauge condition $d^\dagger A=0$ is understood as a five dimensional condition.
This system can be covariantized as follows
\bea
&& d^\dagger \iota_R F=0 ~, \nn\\
&& d^\dagger A=0~, \label{cov-tre-5D}\\
&& F_H^+=0~,\nn
\eea
where $R$ is the vector field associated with the action of $U(1)$ (or a more general toric action). The condition $ F_H^+=0$ is anti self-duality in the horizontal space with respect to $R$, which requires a one form $\kappa$ with the property $\iota_R \kappa=1$. The symbol for the PDEs (\ref{5D-expl-system}) can be written as  follows
\bea
\sigma_{5D}^{tsd} = \left (
\begin{array}{ccccc}
	-{\bf{p}} \cdot {\bf{p}} & p_1 p_0 & p_2 p_0  & p_3 p_0 & p_4 p_0 \\
	p_0 &  p_1 & p_2 & p_3 & p_4 \\
	0& -p_2 & p_1 & - p_4 &  p_3 \\
	0& -p_3 &  p_4 & p_1 & - p_2 \\
	0& -p_4 & - p_3 &  p_2 & p_1
\end{array}\right  )~,\label{matrix5D-non-red}
\eea
where ${\bf{p}} \cdot {\bf{p}} = p_1^2 + p_2^2 + p_3^2 + p_4^2$.  This is the symbol of a transversely elliptic operator with respect to zeroth direction (setting $p_0=0$).
Thus on 5D manifolds with a torus action the system of PDEs (\ref{cov-tre-5D}) gives rise to a transversely elliptic problem and these PDEs are central for for the treatment of 5D supersymmetric Yang-Mills theory \cite{Qiu:2016dyj}.

However, here our main interest is in 4D gauge theory systems. Hence we discuss the reduction of the 5D system (\ref{cov-tre-5D}) down to 4D.
There is always a trivial reduction along $R$ (zeroth direction in our notations) which will bring us back to the elliptic problem in 4D.
We are interested in dimensional reduction along a different direction.  Let us assume that we have at least a $T^2$ action on the five manifold and that it is generated by $\partial_0$ and $\partial_1$.  We can introduce another basis, $\tilde{\partial}_0$
and $\tilde{\partial}_1$ such that
\bea
\partial_0 = \cos \omega ~ \tilde{\partial_0} + \sin \omega~\tilde{\partial}_1~, \nn \\
\partial_1 = - \sin \omega ~ \tilde{\partial_0} + \cos \omega~\tilde{\partial}_1~, \nn
\eea
where $\omega$ is some function invariant under the torus actions. We stress again that our discussion is local, at the level of symbols and
the idea is  to perform the reduction along $\tilde{\partial_0}$ by requiring all fields to be independent of this direction. Almost by definition, the resulting system of PDEs will be transversally elliptic with respect to the remaining $U(1)$ action. At the level of the symbol (\ref{matrix5D-non-red})  we just have to make a substitution $p_0= \sin \omega~ \tilde{p}_1  \equiv s_\omega \tilde{p}_1$ and $p_1= \cos \omega~ \tilde{p}_1 \equiv c_\omega \tilde{p}_1$ with $\tilde{p}_0 =0$
\bea
\sigma_{4D}^{tsd} = \left (
\begin{array}{ccccc}
	- c_\omega^2 \tilde{p}_1^2 - \vec{p} \cdot \vec{p} & c_\omega s_\omega \tilde{p}^2_1  & s_\omega p_2 \tilde{p}_1  & s_\omega p_3 \tilde{p}_1 & s_\omega p_4 \tilde{p}_1 \\
	s_\omega \tilde{p}_1 & c_\omega \tilde{p}_1 & p_2 & p_3 & p_4 \\
	0& -p_2 & c_\omega \tilde{p}_1& - p_4 &  p_3 \\
	0& -p_3 &  p_4 &  c_\omega \tilde{p}_1 & - p_2 \\
	0& -p_4 & - p_3 &  p_2 & c_\omega \tilde{p}_1
\end{array}\right  )~,\label{matrix5D-non-red2}
\eea
where $\vec{p} \cdot \vec{p} = p_2^2 +p_3^2 + p_4^2$.
This matrix, however, is written in the wrong basis $(A_0, A_1, A_2, A_3, A_4)$. Since we perform the reduction along $\tilde{\partial}_0$, the 4D boson $\Phi$ is defined as the $\tilde{\partial}_0$-component of 5D gauge field. Thus we have the following definition
\bea
A_0 = \cos \omega~ \Phi + \sin \omega ~ \tilde{A}_1~,\\
A_1 = - \sin \omega~ \Phi + \cos \omega~ \tilde{A}_1~,
\eea
and in the new basis $(\Phi, \tilde{A}_1, A_2, A_3, A_4)$ the symbol (\ref{matrix5D-non-red}) takes the form
\bea
\tilde{\sigma}_{4D}^{tsd} = \left (
\begin{array}{ccccc}
	- c_\omega ( \tilde{p}_1^2 + \vec{p} \cdot \vec{p}) & - s_\omega \vec{p} \cdot \vec{p}  & s_\omega p_2 \tilde{p}_1  & s_\omega p_3 \tilde{p}_1 & s_\omega p_4 \tilde{p}_1 \\
	0 &   \tilde{p}_1 & p_2 & p_3 & p_4 \\
	s_\omega p_2 & -c_\omega p_2 & c_\omega \tilde{p}_1& - p_4 &  p_3 \\
	s_\omega p_3 & -c_\omega p_3 &  p_4 &  c_\omega \tilde{p}_1 & - p_2 \\
	s_\omega p_4 & -c_\omega p_4 & - p_3 &  p_2 & c_\omega \tilde{p}_1
\end{array}\right  )~.\label{matrix5D-non-red-rot}
\eea
This is exactly the same symbol as in Pestun's analysis on $S^4$ (see equation (4.32) in \cite{Pestun:2007rz}, up to an overall minus in first row) with $\omega$ being identified
with the $\theta$-angle on $S^4$. In our more general story $\omega$ is an effective angle in 5D which controls the relative direction between $R$ and $\tilde{\partial}_0$.
In 4D we have a fixed point when $\sin \omega =0$ and there the symbol (\ref{matrix5D-non-red-rot}) will correspond to the symbol of either the self-dual connection or of the anti self-dual connection problem, depending on the sign of $\cos \omega$ at the fixed point.

All these considerations can be reformulated in terms of a transversely elliptic complex. The 5D language appears to be a natural way to repack the 4D expressions.  It is possible, however, to formulate the corresponding transversally elliptic problem intrinsically in 4D terms.  Let us assume that there is a $U(1)$-action on the four manifold and that the corresponding vector field is $v$. Then one can write down the following PDEs for the gauge potential $A$ and a scalar $\Phi$
(in the context of ${\cal N}=2$ the scalar $\Phi$ is identified with the appropriate combination of the scalars in supersymmetric gauge theory),
\bea
&& d^\dagger \Big (\iota_v F - d(\cos \omega ~\Phi) \Big ) =0 ~, \nn\\
&& d^\dagger A=0~, \label{cov-tre-4D}\label{tran-ell-eqs}\\
&& P_{\omega}^+ \Big ( F + \iota_v (\star d\Phi) \Big )=0~,\nn
\eea
where the projector   $P_{\omega}^+$  is defined in \eqref{projector} and the function $\omega$ is invariant under $v$ and has a specific behavior around the fixed points (see subsection \ref{s:subs-projector} for further details).  Here for the sake of clarity we assume that the metric is such that $||v||^2 = \sin^2 \omega$.
Up to lower derivative terms these PDEs can be obtained by the dimensional reduction of equations (\ref{cov-tre-5D}).  However, they can also be defined intrinsically in 4D terms.
In order to calculate the symbol it is convenient to make a choice of coordinates adapted to the isometry, namely $v= \sin \omega ~p_1$ (by construction $\sin \omega =0$ at the fixed point).
Then  using the form of $P_{\omega}^+$ in these coordinates we obtain the symbol of (\ref{cov-tre-4D}) which is exactly the same as in the expression (\ref{matrix5D-non-red-rot}).
For a more general metric (that is dropping the condition $||v||^2 = \sin^2 \omega$) the first and third equations in (\ref{cov-tre-4D}) should be slightly modified by some additional scaling factors in order to obtain the same symbol (\ref{matrix5D-non-red-rot}).

The equations (\ref{tran-ell-eqs}) can be encoded into the transversely elliptic complex
\bea
\Omega^0 (M_4) ~\xrightarrow{d}~\Omega^1 (M_4) \oplus \Omega^0(M_4)~\xrightarrow{\tilde{D}}~P^+_\omega\Omega^{2} (M_4)\oplus \Omega^0 (M_4)~,
\label{complex-explicit}
\eea
which we denote as $(E^{\sbullet},\gdh)$. Here the explicit form of the operator $\tilde{D}$ can be read off from the first and third equations in (\ref{tran-ell-eqs}) and it is given by
 \bea
 \tilde{D} = \left ( \begin{array} {cc}
     P_\omega^+ d  & P_\omega^+ \iota_v \star d \\
        d^\dagger \iota_v d  &  - d^\dagger d \cos \omega
        \end{array} \right )\label{expl-form-tildeD}
 \eea
written in the basis $\Omega^1 (M_4) \oplus \Omega^0(M_4)$. Remember that here we expand around the zero connection for the abelian problem.
It is straightforward to generalize the equations (\ref{cov-tre-5D}),  (\ref{cov-tre-4D}) and (\ref{complex-explicit}) to the non-abelian case  and around a non-zero connection.
However, this will not affect our discussion of the symbols.

\section{Full cohomological complex} \label{app2}

In subsection \ref{subs:cohom-complex} we defined the cohomological supersymmetry transformation (\ref{complextransf}) on the set of fields $(A, \Psi, \phi, \varphi, \eta, \chi, H)$.
In the full theory we have to include more fields related to the gauge fixing and combine appropriately the cohomological supersymmetry  (\ref{complextransf}) with the BRST transformations.
Let us introduce the ghost $c$ which is an odd zero form in the adjoint representation, the anti-ghost $\bar{c}$ which is an odd zero form in the adjoint representation and a Lagrangian multiplier $b$ which is an even zero form in the adjoint representation.
The full cohomological transformations are defined as follows
\be \label{full-cohom-tran}
\begin{aligned}
	&\delta A =  i \Psi + d_A c~ , \\
	& \delta \varphi = i\eta  + i[c,\varphi]~, \\
	& \delta \chi = H + i \{ c, \chi \} ~ ,  \\
	& \delta c =  \phi + i \iota_v A +i \{c,c\}/2 ~, \\
	& \delta \bar{c} =b  + i \{c , \bar{c}\}~ ,
\end{aligned}
\qquad
\begin{aligned}
	& \delta \Psi =  \iota_{v} F  +  i d_A \phi  +  i \{ c, \Psi \}  ~, \\
	& \delta \eta = {\cal L}_{v}^A \varphi - [\phi,  \varphi ] + i \{ c, \eta \} ,\\
	&  \delta H = i {\cal L}_{v}^A \chi - i [\phi,\chi ] +i [ c, H ] ~,  \\
	& \delta \phi =  \iota_{v} \Psi  + i [c, \phi ]~,   \\
	& \delta b = i {\cal L}^A_v \bar c - i [ \phi, \bar{c}] + i [c, b] ~ ,
\end{aligned}
\ee
where we assume the same conventions as in subsection \ref{subs:cohom-complex} and on all fields we have $\delta^2 = i {\cal L}_v$. 
 Actually we have to choose a background connection around which we expand  and in above transformations we expand around zero connection.
These cohomological transformations and their linearization have an intricate structure which is discussed in detail in \cite{Fest-2018}.

For localization  
 we need to treat zero modes more systematically. Following  \cite{Pestun:2007rz} we introduce the zero mode sector $(a_0, \bar{a}_0, c_0, \bar{c}_0, b_0)$.
  In the transformation (\ref{full-cohom-tran}) we modify the transformation for the ghost $c$ as follows
\be
\begin{aligned}
	\delta c = a_0+ \phi + i \iota_v A +\frac{i}{2} \{c,c\}~ ,
\end{aligned}
\ee
 and in addition we define the following transformations for zero modes
\be  \label{tran-zero-modes}
\begin{aligned}
	&\delta  \bar{a}_0 = \bar{c}_0 ~, \\
	&\delta b_0 = c_0 ~ , \\
	&\delta a_0 = 0~ .
\end{aligned}
\qquad
\begin{aligned}
	&\delta \bar{c}_0 = i[a_0,\bar{a}_0 ]~ ,  \\
	&  \delta c_0 = i[a_0,b_0] ~ ,  \\
	& \
\end{aligned}
\ee
The modified transformations satisfy the following algebra
\bea \label{delta_2_full}
\delta^2 = i {\cal L}_v + G_{a_0} ,
\eea
on all fields.

\section{Supergravity background \label{appE}}

Here we collect together explicit expressions for the background supergravity fields necessary to solve the generalized Killing spinor equations as described in section\ref{sec:killingspinorsol}. These formulae make use of the spinor bilinears  $\Theta_{i j}~,\widetilde \Theta_{ij}$ and $v^{\mu}_{i j}$ that are as defined in \eqref{bilspi}.

The SU(2)$_R$ connection $(V_\mu)_{i j}$ and two form $W_{\mu\nu}$ are given by
\begin{align}
\label{solTVauxhh}
W_{\mu\nu} =& {i \over s+\tilde s} (\partial_\mu v_\nu-\partial_\nu v_\mu)-{2i \over (s+\tilde s)^2} {\epsilon_{\mu\nu\rho}}^{\lambda}v^\rho \partial_\lambda(s-\tilde s)-{4 \over s+\tilde s}{\epsilon_{\mu\nu\rho}}^\lambda v^\rho G_\lambda+\cr
& +{s-\tilde s\over (s+\tilde s)^2}{\epsilon_{\mu\nu\rho}}^\lambda v^\rho b_\lambda +{1\over s+\tilde s}(v_{\mu} b_{\nu} -v_{\nu}b_{\mu})~,\cr
(V_\mu)_{i j}=&{4\over s+\tilde s}\left(\zeta_{(i}\nabla_\mu \zeta_{j)}+\bar \chi_{(i}\nabla_\mu \bar \chi_{j)}\right)+{4\over s+\tilde s} \left(2 i G_\nu-{\partial_\nu(s-\tilde s)\over (s+\tilde s)}\right){(\Theta_{i j}-\widetilde \Theta_{i j})^\nu}_\mu+\cr&+ {4i\over (s+\tilde s)^2}b_\nu {( \tilde s\, \Theta_{i j}+s\,\widetilde  \Theta_{i j} )^\nu}_\mu~,
\end{align}

The graviphoton field strength and the scalars $S_{ij}$ take the form
\begin{align}
\label{solS}
& {\mathcal F}_{\mu\nu}=i \partial_{\mu}\Big({s+\tilde s-K\over s \tilde s} v_\nu\Big)-i\partial_{\nu}\Big({s+\tilde s-K\over s \tilde s} v_\mu\Big)~,\cr
& S_{i j}={8 i \over (s+\tilde s)^2} (\Theta_{i j}+\widetilde \Theta_{ij})^{\mu\nu}\partial_\mu v_\nu- 2i{ s+\tilde s-K\over (s\, \tilde s)^2} (\tilde s\,\Theta_{i j}+s\, \widetilde \Theta_{ij})^{\mu\nu}\partial_\mu v_\nu+\cr
&\quad \,\,+{2\over s+\tilde s} \Big({4}G_\mu-{s-\tilde s\over s+\tilde s} b_\mu-{i\over 2} {s-\tilde s\over s \tilde s}\partial_\mu(s+\tilde s)\Big) v^{\mu}_{i j}~.
\end{align}
Finally the scalar combination $R/6-N$ is given by
\begin{align}
\label{solN}
\Big({R\over 6}-N\Big)=&~{s-\tilde s\over(s+\tilde s )^2 }\Box(s-\tilde s)+{ \partial_{[\mu} v_{\nu]}\partial^{[\mu}v^{\nu]}\over (s+\tilde s)^2}-{1\over 2}{s-\tilde s \over (s+\tilde s)^3} \epsilon^{\mu\nu\rho\lambda}(\partial_\mu v_\nu)(\partial_\rho v_{\lambda})+\cr
&-{4 \epsilon^{\mu\nu\rho\lambda}\over (s+\tilde s)^3} v_\mu(\partial_\nu v_\rho)\partial_\lambda(s-\tilde s)+{2 s \tilde s\over (s+\tilde s)^4}\partial_\mu(s-\tilde s)\partial^\mu (s-\tilde s)+\cr
&-2{s -\tilde s\over (s+\tilde s)^3}\partial_\mu(s-\tilde s)\partial^\mu (s+\tilde s)-{2i s \tilde s\over (s+\tilde s)^2} \nabla^\mu b_\mu+{2(s \tilde s)^2\over (s+\tilde s)^4} b_\mu b^\mu+\cr
&-i{s-\tilde s\over (s+\tilde s)^3} \epsilon^{\mu\nu\rho\lambda}(\partial_\mu v_\nu)v_\rho b_\lambda+3i{s-\tilde s\over (s+\tilde s)^3} b^\mu (\tilde s \partial_\mu s-s\partial_\mu \tilde s)+\cr &
+2i{ s^2+ \tilde s^2 \over (s+\tilde s)^4}b^\mu \partial_\mu(s \tilde s)-2i{s-\tilde s \over s+\tilde s}\nabla^\mu G_\mu+ 4{s^2+{\tilde s}^2\over (s+\tilde s )^2} G^\mu G_\mu+{8 (v^\mu G_\mu)^2\over (s+\tilde s)^2}+\cr &
+4{s\tilde s} {s-\tilde s \over(s+\tilde s)^3}G_\mu b^\mu+{ 4i\epsilon^{\mu\nu\rho\lambda} \over(s+\tilde s)^2} (\partial_\mu v_\nu)v_\rho G_\lambda+4i{s-\tilde s \over (s+\tilde s)^3} G^\mu\partial_\mu(s \tilde s)~.
\end{align}

\providecommand{\href}[2]{#2}\begingroup\raggedright

\bibliographystyle{utphys}
\bibliography{4Dgaugetheory}{}

\providecommand{\href}[2]{#2}\begingroup\raggedright

\endgroup

\end{document}